\documentclass[accepted]{melba}

\usepackage{mwe} 

\usepackage{amsmath,amsfonts}

\usepackage{algorithm}
\usepackage{algorithmic}
\usepackage{booktabs}
\usepackage{url}
\usepackage[algo2e, ruled]{algorithm2e}

\melbaid{2025:031}  
\doi{https://doi.org/10.59275/j.melba.2025-79a2}
\melbaauthors{Chen, Konz, Gu, Dong, Chen, Li, Lee and Mazurowski}  
\email{yuwen.chen@duke.edu}
\volume{3}
\firstpageno{711}  
\melbayear{2025}  
\datesubmitted{2025-07-10}  
\datepublished{2025-11-25}  

\ShortHeadings{ContourDiff}{Chen, Konz, Gu, Dong, Chen, Li, Lee and Mazurowski}

\title{ContourDiff: Unpaired Medical Image Translation with Structural Consistency}

\author{
	\firstname Yuwen \surname Chen\aff{1} \orcid{0009-0003-9826-1013},
	\firstname Nicholas \surname Konz \aff{1},
        \firstname Hanxue \surname Gu \aff{1},
        \firstname Haoyu \surname Dong \aff{1},
        \firstname Yaqian \surname Chen \aff{1},
        \firstname Lin \surname Li \aff{2},
        \firstname Jisoo \surname Lee \aff{3},
        \firstname Maciej A. \surname Mazurowski \aff{1,2,3,4}
}
\affiliations{
	\num 1 \addr Department of Electrical and Computer Engineering, Duke University, Durham, NC 27708 \\
        \num 2 \addr Department of Biostatistics \& Bioinformatics, Duke University, Durham, NC 27708 \\
	\num 3 \addr Department of Radiology, Duke University, Durham, NC 27708 \\
        \num 4 \addr Department of Computer Science, Duke University, Durham, NC, 27708
}

\abstract{
	Accurately translating medical images between different modalities, such as Computed Tomography (CT) to Magnetic Resonance Imaging (MRI), has numerous downstream clinical and machine learning applications. While several methods have been proposed to achieve this, they often prioritize perceptual quality with respect to output domain features over preserving anatomical fidelity. However, maintaining anatomy during translation is essential for many tasks, e.g., when leveraging masks from the input domain to develop a segmentation model with images translated to the output domain. To address these challenges, we propose ContourDiff with Spatially Coherent Guided Diffusion (SCGD), a novel framework that leverages domain-invariant anatomical contour representations of images. These representations are simple to extract from images, yet form precise spatial constraints on their anatomical content. We introduce a diffusion model that converts contour representations of images from arbitrary input domains into images in the output domain of interest. By applying the contour as a constraint at every diffusion sampling step, we ensure the preservation of anatomical content. We evaluate our method on challenging lumbar spine and hip-and-thigh CT-to-MRI translation tasks, via (1) the performance of segmentation models trained on translated images applied to real MRIs, and (2) the foreground FID and KID of translated images with respect to real MRIs. Our method outperforms other unpaired image translation methods by a significant margin across almost all metrics and scenarios. Moreover, it achieves this without the need to access any input domain information during training and we further verify its zero-shot capability, showing that a model trained on one anatomical region can be directly applied to unseen regions without retraining. Our code is available at~\url{https://github.com/mazurowski-lab/ContourDiff}.}

\keywords{Unpaired Image Translation, Medical Image Segmentation, Diffusion Model}

\begin{document}

\twocolumn[\maketitle]

\section{Introduction}
\label{sec:introduction}
	\enluminure{U}{npaired} image-to-image (I2I) translation---the task of translating images from some input domains to an output domain with only unpaired data for training \cite{cyclegan}---offers extensive applications in medical image analysis \cite{armanious2020medgan,durrer2024diffusion,beizaee2023harmonizing,wang2024mutual,modanwal2020mri,yang2019unsupervised,liu2021style,zhang2018translating}. A significant use case is facilitating segmentation across different imaging modalities (e.g., CT and MRI) \cite{chen2023deep}, for anatomical locations such as brain \cite{li2020magnetic}, abdomen \cite{synsegnet}, and pelvis \cite{rossi2021comparison}. This approach is especially beneficial given the significant time and labor involved in annotating images for each modality independently. Through direct image translation between modalities, annotations from one modality can be reused in another, reducing manual effort. However, achieving this requires strict anatomical consistency in translation.

    \begin{figure}[!ht]
        \centering
        \includegraphics[width=0.4\textwidth]{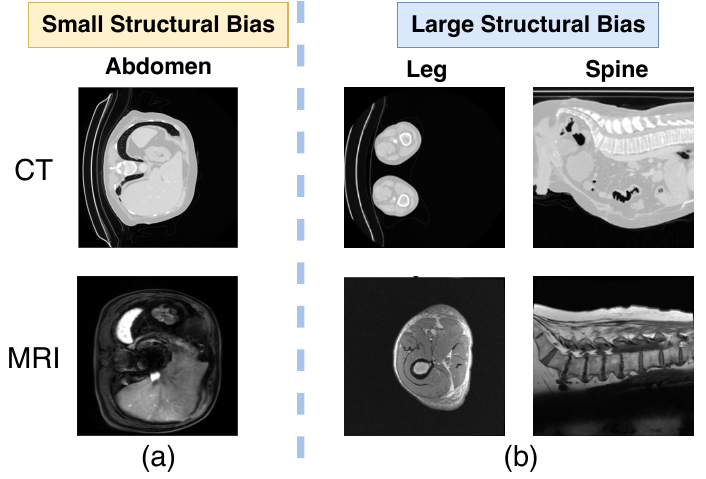}
        \caption{\textbf{Structural biases between CT and MRI modalities in certain anatomical regions:} minor for the abdominal region from axial view (a), but severe for the leg from axial view and spinal regions from sagittal view (b).}
        \label{fig:teaser}
    \end{figure}

    Ensuring anatomical consistency in unpaired I2I translation is challenging, particularly when the input and output domains exhibit a substantial \textit{structural bias}---i.e., a consistent difference in anatomical structure and shape between domains. An example of this is the drastic visual difference between CT and MRI from different protocols for leg and spinal regions as captured in standard exams (see Fig. \ref{fig:teaser} and \ref{fig:translation}), where typically CT images display two legs while MRI scans only show one, and CT images capture entire the abdominal body while MRI focuses on the lumbar area, respectively. Traditional translation models tend to internalize this structural bias, resulting in them applying drastic anatomical transformations during translation in order to align with the typical structure seen in the output domain, resulting in a misalignment between translated images and their corresponding input segmentation masks, potentially leading to unreliable segmentation models trained this data. 
    
    One group of methods for unpaired I2I translation in medical imaging is based on Generative Adversarial Networks (GANs) \cite{goodfellow2020generative} such as Cycle-consistent Adversarial Network (CycleGAN) \cite{cyclegan,armanious2019unsupervised,chen2023deep,maskgan,zhou2023gan}. These methods maintain the consistency between the images from input and output domains by leveraging cycle consistency loss, minimizing information loss during bidirectional translation \cite{cyclegan}. However, such cycle-consistent supervision does not provide a direct and interpretable constraint on preserving anatomical structures between modalities. Indeed, CycleGAN and its variants may yield undesirable results when substantial misalignment exists between modalities \cite{maskgan}.
    
    Recently, several conditional diffusion models have been introduced for image translation tasks, both in natural images \cite{batzolis2021conditional,rombach2022high,li2023bbdm,unsb} and medical imaging \cite{li2023zero,ozbey2023unsupervised,kim2024adaptive}. However, some of these methods are constrained to paired data or aligning features in domains that are difficult to interpret for unpaired data, such as latent or frequency domains. 
    
    To preserve anatomical structures using pixel-level constraints, inspired by previous works in spatially-conditioned diffusion models \cite{konz2024anatomically,controlnet,rombach2022high}, we propose a diffusion model for image translation, ``\textbf{ContourDiff}'' \footnote{Code: \url{https://github.com/mazurowski-lab/ContourDiff}}, that uses domain-invariant anatomical contour representations of images to guide the translation process, which enforces precise anatomical consistency even between modalities with severe structural biases. This model also has the added benefit of \textbf{allowing zero-shot learning}: it solely requires a set of unlabeled output domain images for training, unlike most unpaired translation models. As such, it can potentially translate images from arbitrary unseen domains at inference (see Section \ref{sec:t2tot1}), which can be advantageous for medical image harmonization across multiple imaging modalities. We evaluate our method on CT to MRI translation for sagittal-view lumbar spine and axial-view hip-and-thigh body regions, which both possess severe structural biases (Fig. \ref{fig:teaser} and \ref{fig:translation}). 
    In addition to utilizing standard unpaired image generation quality metrics like FID and KID, we evaluate the anatomical consistency of our translation model by training a segmentation model on CT images translated to MRI given their original masks, and evaluating it for real MRI segmentation. Our main contributions include:
    \begin{enumerate}
        \item We propose ContourDiff, a novel diffusion-based framework for unpaired image-to-image translation which allows zero-shot learning.
        \item We introduce Spatially Coherent Guided Diffusion (SCGD) to enforces spatial consistency within a volume by providing context information from adjacent slices.
        \item Our method significantly outperforms existing unpaired I2I models, including GAN-based and diffusion-based methods, in segmentation performance over all test datasets, despite the fact that it requires no input domain information for training, unlike the competing methods.
        \item Our method achieves the best performance compared to existing I2I models in terms of foreground FID and KID across almost all situations.
        \item We demonstrate the zero-shot capability of ContourDiff by translating additional input-domain modalities to the output domain without any model retraining.
    \end{enumerate}

\section{Related Works}
\label{sec:related_works}
    \subsection{Image-to-Image Translation}
    Image-to-image translation aims to learn a mapping to transform images from one domain to another while preserving essential structural details. Several GAN-based frameworks, including Pix2Pix \cite{isola2017image} and its variants \cite{wang2018high}, have been developed as supervised learning methods for paired image-to-image translation. GAN-based models are also widely used in unpaired translation, with CycleGAN \cite{cyclegan} introducing cycle-consistency loss to allow translation between unpaired datasets. MUNIT \cite{munit} enables multi-modal outputs to generate diverse outputs given images from input domains. GcGAN \cite{gcgan} incorporates geometric-consistency constraints to preserve the geometric information across domains. To reduce the training time, CUT \cite{cut} leverages contrastive learning to align corresponding patches between domains in feature space, instead of using entire images. Despite the success, GAN-based techniques often face challenges like training instabilities and mode collapse problems \cite{li2023bbdm}. More recently, diffusion-based translation frameworks have emerged as a promising alternative, providing competitive performance in both paired \cite{li2023bbdm} and unpaired \cite{unsb} image translation tasks.
    
    Image-to-image translation specialized for medical imaging aims to convert images between modalities (e.g., CT to MRI) to generate synthetic data and improve diagnostic capabilities. However, acquiring labeled and paired medical images is both challenging and expensive \cite{chen2023moco}, which exacerbates the challenge of preserving anatomical structures---an essential aspect in medical image translation. To address this issue, several GAN-based frameworks have been developed for \textit{unpaired} medical image translation
    \cite{armanious2019unsupervised,uzunova2020memory,kong2021breaking}. Recently, diffusion models have gained popularity in this domain. For instance, SynDiff \cite{ozbey2023unsupervised} incorporates the adversarial diffusion modeling to achieve unsupervised medical image translation. However, these methods rely on adversarial training to align features, lacking strict and interpretable constrains on the detailed anatomical structures during translation.
    
    \subsection{Diffusion Models}
    Denoising Diffusion Probabilistic Models (DDPM) \cite{ho2020denoising}, or just \textit{diffusion models}, have recently gained significant attention for their remarkable performance in generative modeling across both natural \cite{croitoru2023diffusion,muller2023multimodal} and medical imaging tasks \cite{rombach2022high,konz2024anatomically}. Different from GAN-based models, diffusion models generate high-quality images with progressive denoising steps, starting from random noise and gradually refining it into a coherent image. Conditional diffusion models extend this approach by incorporating additional conditions, such as texts and images, into the training objectives and model input. For instance, Konz et al. \cite{konz2024anatomically} guided the generation process of medical images with pixel-level masks at each denoising step to ensure strict spatial control over the output. Latent Diffusion Models (LDMs) \cite{rombach2022high} on the other hand shift the diffusion process to a lower-dimensional latent space rather than operating in pixel space for better computational scaling to large images; however, working in this latent space requires a loss of fine detail in the images which the model is conditioned on (in our case, the anatomical contour map) due to downsampling, so our approach remains in image space. Conditional diffusion models have also been explored for other image-to-image tasks, including inpainting \cite{rombach2022high,corneanu2024latentpaint}, super-resolution \cite{saharia2022image,gao2023implicit} and semantic segmentation \cite{tan2022semantic,baranchuk2021label}.

\section{Methods}
\label{sec:methods}

    \begin{figure*}[!ht]
        \centering
        \includegraphics[width=1\textwidth]{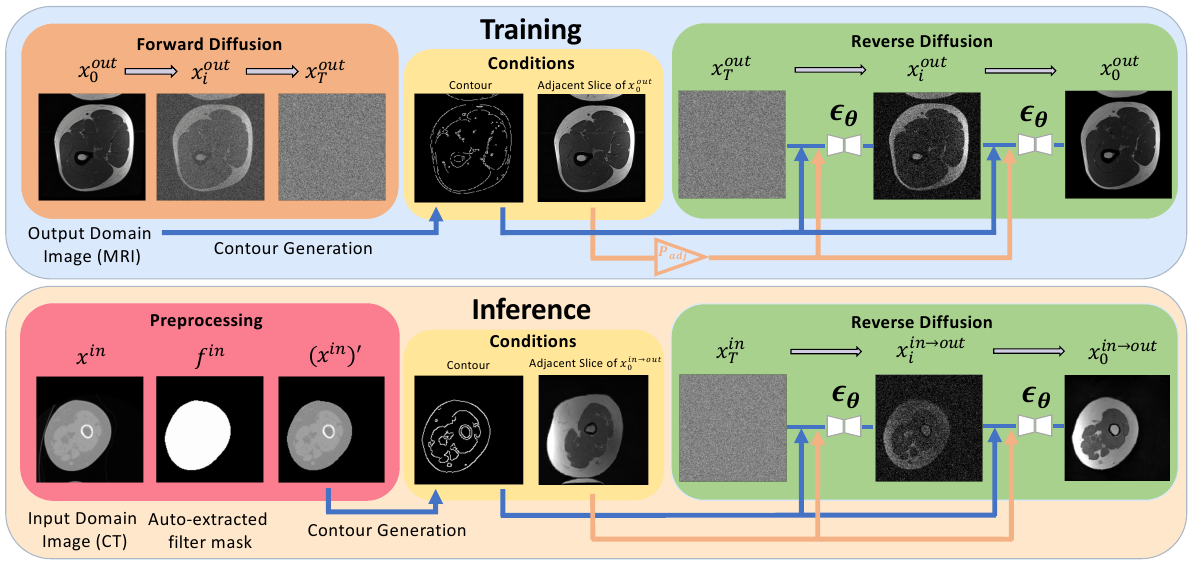}
        \caption{\textbf{Overview of ContourDiff.} Top is the training process of ContourDiff. The denoising model $\epsilon_{\theta}$ is trained on output domain images, conditioning on their anatomical contours and on an adjacent slice with probability $P_{adj}$. Bottom is the inference process of ContourDiff. The model generates input domain images in the appearance of the output domain given input domain contours and previously generated adjacent slices.}
        \label{fig:figure_pipeline}
    \end{figure*}
    
    \subsection{Problem Definition} 
    In unpaired image translation, only unpaired datasets of input and output domain examples are available for training. Our method is even more general in that it accomplishes \textit{zero-shot} image translation, where only an unlabeled dataset of $N_{out}$ output domain examples $[x^{out}]_n$ ($n=1,\ldots,N_{out}$) are available to train on. The goal is then to use the trained model at inference to translate unseen input domain data $[x^{in}]_n$ to the output domain. In our case, we aim to translate CT images to the MRI domain, for usage with MRI-trained segmentation models. To do so, we propose a novel diffusion-based image translation framework based on domain-invariant anatomical contours of images.
    
    \subsection{Adding Contour Guidance to Diffusion Models}
    \subsubsection{Diffusion Models}
    Denoising diffusion probabilistic models \cite{ho2020denoising} are generative models that learn to reverse a gradual process of adding noise to an image over many time steps $t=0,\ldots, T$. New images can be generated by starting with a (Gaussian) noise sample $x_T$ and iteratively applying the model to obtain $x_{t-1}$ from $x_t$ for $t=T,\ldots, 0$ until an image $x_0$ is recovered.
    
    In practice, the neural network itself $\epsilon_\theta (x_t,t)$ is an I2I architecture (e.g., a UNet \cite{unet}) that is trained to predict the noise $\epsilon$ added to an image $x_0$ at various timesteps $t$. The training objective is to optimize the Evidence Lower Bound (ELBO). The loss can be simply described as \cite{nichol2021improved}:
    \begin{equation}
        L=\mathbb{E}_{x_0,t,\epsilon} \left [||\epsilon - \epsilon_\theta(x_t,t)||^2\right]
    \end{equation}
    where $\theta$ is the model parameters.
    
    Unlike unconditional DDPMs, many conditional diffusion models \cite{rombach2022high,li2023bbdm,konz2024anatomically} directly integrate the conditions $y$ (e.g., images and texts) into the training objective: 
    \begin{equation}
    \label{eq:uncond_loss}
        L=\mathbb{E}_{(x_0,y),t,\epsilon} \left [||\epsilon - \epsilon_\theta(x_t,t|y)||^2\right],
    \end{equation}
    which allows the model to leverage external information to guide the generation process.
    
    Denoising Diffusion Implicit Models (DDIMs) \cite{song2020denoising} employ a deterministic, non-Markovian sampling process, allowing for faster sample generation without noticeable compromises for image fidelity.
    
    \subsubsection{Contour-guided Diffusion Models}
    For standard unconditional diffusion models, it is unclear how to constrain the semantics/anatomy of generated images. To address this, we propose to utilize \textit{contour} representations of images to provide guidance in generating the image. While training the model, we use the Canny edge detection filter \cite{canny1986computational} to extract the contour representation $c$ of each training image $x_0$, similar as that in \cite{rombach2022high}, and concatenate it with the network input at every denoising step, a practice similar to \cite{konz2024anatomically,controlnet}. This modifies the network in Eq. \ref{eq:uncond_loss} to become $\epsilon_\theta(x_t, t|c)$ and the diffusion training objective to become
    \begin{equation}
    \label{eq:loss}
        L=\mathbb{E}_{(x_0,c),t,\epsilon} \left [||\epsilon - \epsilon_\theta(x_t,t|c)||^2\right],
    \end{equation}
    where $(x_0,c)$ is a training set image and its accompanying contour. We perform this in image space in order to ensure that the denoised image precisely follows the contour guidance pixel-to-pixel (as in \cite{konz2024anatomically}), which may be lost if diffusion is performed within a latent space \cite{rombach2022high}. 
    
    \subsection{Contour-guided image translation}
    \begin{figure*}[t]
        \centering
        \includegraphics[width=1\textwidth]{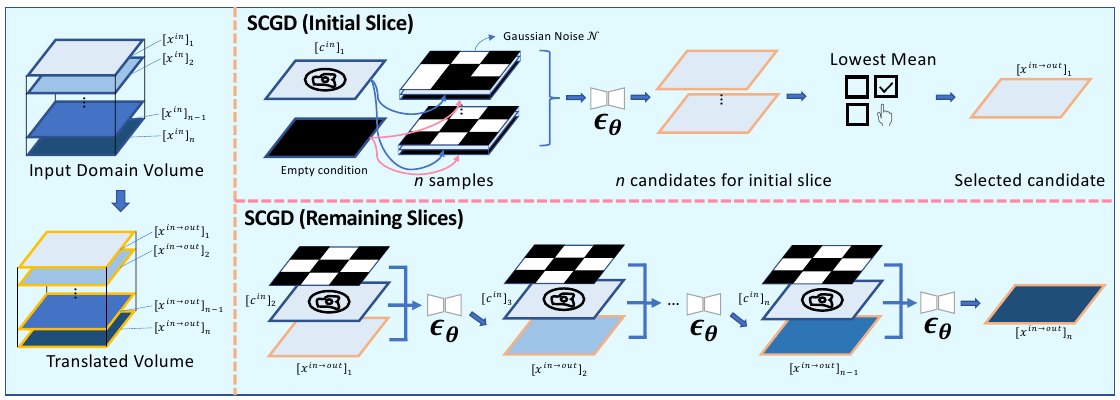}
        \caption{\textbf{Spatially Coherent Guided Diffusion (SCGD).} For each input domain volume, SCGD first translates the initial slice by generating $n$ candidates with setting $C_{adj}$ to an empty map and selecting the optimal one according to a specified criterion (e.g., lowest mean intensity). Then, every subsequent slice is synthesied by conditioning on its anatomical contours and the previously translated slice. Input domain slices are bordered in blue and output domain slices are bordered in orange.}
        \label{fig:scgd_pipeline}
    \end{figure*}
    \subsubsection{Overall Translation Process}
    One important feature of contours is that they can be viewed as domain-invariant yet anatomy-preserving representations of images. This allows for a contour-guided diffusion model trained in some output domain to serve as a zero-shot image translation method, as follows. 
    
    First, we train a contour-guided diffusion model on output domain images with accompanying contours $([x^{out}]_n,[c^{out}]_n)$, shown in Algorithm \ref{alg:train} (Note: Algorithm \ref{alg:train} also include constraints from adjacent slices). Next, to translate some \textit{input domain} image $x^{in}$ to the output domain, we extract its contour $c^{in}$ after removing irrelevant backgrounds using $F_{filter}$, and use the output domain-trained model $\epsilon_\theta$ conditioned on $c^{in}$ to generate the image $x^{in \rightarrow out}$. Therefore, $x^{in \rightarrow out}$ maintains the anatomical content of $x^{in}$, while possessing the visual domain characteristics of the output domain. Our translation algorithm is shown in Algorithm \ref{alg:translate}, where $\alpha_t= 1- \beta_t$ with the variance of the additive pre-scheduled noise $\beta_t$, and $\overline{\alpha}_t=\prod_{s=1}^t\alpha_s$.
    
    \subsubsection{Filtering Out Image Artifacts}
    We also apply additional pre-processing to network input images $x$ to filter out non-anatomical features/artifacts (e.g., the motorized table in CT) if necessary, by applying a binary mask $M_{filter}$ as $x\leftarrow M_{filter}\odot x$. $M_{filter}$ is defined by sequentially computing the follow Scikit-Image \cite{van_der_Walt_2014} functions on $x$ \cite{maskgan}: $\texttt{threshold\_multiotsu}$, $\texttt{binary\_erosion}$, $\texttt{remove\_small\_objects}$, and $\texttt{remove\_small\_holes}$.
    
    \subsection{Spatially Coherent Guided Diffusion (SCGD)}
    We introduce Spatially Coherent Guided Diffusion (\textbf{\emph{SCGD}}), a novel framework designed to preserve spatial consistency when translating adjacent slices from 3D volumes (e.g., CT) into an output domain. SCGD jointly leverages anatomical contour information from the current slice and spatial contextual guidance from its neighbors to enforce both spatially coherent and anatomically consistent translations for each volume. To enable such joint conditioning, SCGD introduces an additional input channel for the diffusion model, $C_{adj}$.
    
    \subsubsection{Training}
    During training, each reverse diffusion step is conditioned on the contour of the current slice, $[x^{in}]_i$, together with one adjacent slice. As slice ordering may be indeterminate at inference, we randomly choose either $[x^{in}]_{i+1}$ or $[x^{in}]_{i-1}$ with equal probability to enable bidirectional spatial guidance.
    
    \paragraph{Adjacent Slice Ratio in Training}
    To ensure the model learns contour-based image generation, we incorporate adjacent slices as conditioning inputs with probability $P_{adj}$. For each training datapoint, $C_{adj}$ is set as:
    \begin{equation}
        C_{adj} = \biggl\{^{[x_0^{out}]_{adj} \: \text{, with probability} P_{adj}}_{\mathbf{0}_{2D} \: \text{, with probability } 1 - P_{adj}}
    \end{equation}
    
    where $[x_0^{out}]_{adj}$ is the adjacent output-domain slice and $\mathbf{0}_{2D}$ is the empty condition map. Intuitively, the model should rely more on anatomical contour information from the current input slice than on information from adjacent slices. To enforce this, we set $P_{adj}$ no larger than 0.5. This choice is particularly important for translating the initial slice for each volume at inference, where no adjacent translated slice is available. Moreover, keeping $P_{adj}$ moderate can prevent the accumulation of errors across slices.

    \begin{algorithm}[H]
      \caption{ContourDiff Training Phase} \label{alg:train}
      \KwIn{Output domain training distribution $p(x_0^{out})$.}
        \Repeat{converged}{
          \ \ $x_0^{out}\sim p(x_0^{out})$\\
          $c^{out} = \texttt{Canny}(x_0^{out})$ \\
          \eIf{rand() $\leq P_{adj}$}{
                $[x_{0}^{out}]_{adj} = $ \texttt{Slice adjacent to} \: $x_{0}^{out}$\;
          }{
                $[x_{0}^{out}]_{adj} = \mathbf{0}$\;
          }
          $\epsilon\sim\mathcal{N}(0,I_n)$\\
          $t \sim \mathrm{Uniform}(\{1, \dotsc, T\})$\\
          $x^{out}_t = \sqrt{\bar\alpha_t} x_0^{out} + \sqrt{1-\bar\alpha_t}\epsilon$\\
          Update $\theta$ with $\: \nabla_\theta \left\| \epsilon - \epsilon_\theta(x_t^{out}, t|(c^{out}, [x_{0}^{out}]_{adj}) \right\|^2$\\
        }
    \end{algorithm}
    
    \begin{algorithm}[H]
      \caption{ContourDiff Inference Phase} \label{alg:translate}
      \KwIn{Input domain image $x^{in}$.}
      \KwOut{Translated image $x_0^{in \rightarrow out}$}
        \ \ $c^{in} = \texttt{Canny}(x^{in})$ \\
        $x^{out}_T \sim \mathcal{N}(0, I_n)$\\
        \eIf{Initial Slice}{
            $[x_{0}^{in \rightarrow out}]_{adj} = \mathbf{0}$\;
        }{
            $[x_{0}^{in \rightarrow out}]_{adj} = $ \texttt{Slice adjacent to} \: $x_{0}^{in \rightarrow out}$\;
        }
        \For{$t=T, \ldots, 1$}{
          \ \ $\epsilon \sim \mathcal{N}(0, I_n)$ if $t > 1$, else $\epsilon = 0$\\
          $x^{out}_{t-1} = \frac{1}{\sqrt{\alpha_t}} \\ \times \left( x^{out}_t - \frac{1-\alpha_t}{\sqrt{1-\bar\alpha_t}} \epsilon_\theta(x^{out}_t, t|(c^{in}, [x_{0}^{in \rightarrow out}]_{adj})) \right) \\ + \sigma_t\epsilon$\\
        }
        \textbf{return} $x_0^{in \rightarrow out}$\\
    \end{algorithm}
    
    \subsubsection{Inference}
    As illustrated in the Fig. \ref{fig:scgd_pipeline}, we first translate the initial slice, $[x^{in}]_1$, of a given 3D volume to its output domain version $[x^{in \rightarrow out}]_1$. To obtain a robust starting point, we generate $n$ samples in parallel (we use $n=16$ in our experiments) and select the sample exhibiting the lowest mean intensity, as we empirically observe that instability in stochastic reverse diffusion can produce overly bright backgrounds. Then, the remaining slices within the volume are translated sequentially, conditioning each step on both the contour of the current input slice and on the previous translated slice to preserve anatomical consistency and spatial coherence throughout the volume. To accelerate the translation, we provide a volume-group parallel inference implementation that processes multiple groups in parallel. Specifically, we partition all volumes evenly into groups and process each group concurrently via scripts. The overall pseudocode for the training and inference phases with SCGD is presented in Algorithms \ref{alg:train} and \ref{alg:translate}, respectively.

\section{Experiment}
\label{sec:experiment}
\subsection{Datasets}
    In this paper, we study one of the most common translation scenarios, CT to MRI, based on three datasets: \textbf{TotalSegmentator} public dataset \cite{Wasserthal_2023}, \textbf{SPIDER} lumbar spine (L-SPIDER) public dataset \cite{vandergraaf2023lumbar} and a private in-house dataset. 
    
    For the MRIs used to train the ContourDiff, we collect a private dataset with T1 weighted lumbar spine (L) and hip \& thigh (H\&T) body regions. 40 sagittal lumbar MRI volumes (670 2D slices), and 10 axial MRI volumes from thigh and hip (404 2D slices) are selected. Correspondingly, we obtain 54 sagittal (2,333 2D slices) and 29 axial (4,937 2D slices) CT volumes from the TotalSegmentator \cite{Wasserthal_2023} in L and H\&T, respectively. 
    
    For downstream bone segmentation task, we further randomly split the two CT sets by patients (43:11 for L and 23:6 for H\&T) for training and validation. We evaluate the segmentation performance on held-out annotated MRI sets (10 L volumes including 158 2D slices, 12 H\&T volumes including 426 2D slices). In addition, to study the generalization ability of our method, we test the lumbar segmentation model on 40 volumes (731 2D slices) from L-SPIDER \cite{vandergraaf2023lumbar} \footnote{We crop the slices to exclude the sacrum, as it is not annotated.}. Moreover, we collect an additional held-out annotated MRI dataset to train the segmentation model directly on real output-domain images (352:80 2D slices for L; 990:305 2D slices for H\&T), which serves as an upper bound (UB) for performance.
    
    \subsection{Evaluation Metrics}
    We quantitatively evaluate translation performance by first training segmentation models on translated images with input domain (CT) masks and testing on real output domain (MRI) images. We adopt commonly-used metrics, Dice Coefficient (DSC) and average symmetric surface distance (ASSD), both evaluated on 3D volumetric segmentation. Given the predicted mask $A$ and the ground truth mask $B$, the two metrics are defined as:
    \begin{equation}
        DSC(A, B) = \frac{2|A \cap B|}{|A| + |B|}
    \end{equation}
    DSC measures the overlap between $A$ and $B$, ranging from 0 and 1. Higher DSC represents better segmentation performance.
    \begin{equation}
        ASSD(A, B) = \frac{\sum_{p \in S_A} d(p, S_B) + \sum_{q \in S_B} d(q, S_A)}{|S_A| + |S_B|}
    \end{equation}
    where $S_A$ and $S_B$ are the sets of surface points of mask $A$ and $B$. $d(p, S_B)$ and $d(q, S_A)$ are the shortest Euclidean distances from a surface point in $S_A$ and $S_B$ to the nearest point in $S_B$ and $S_A$, respectively. ASSD penalizes more on boundary errors relative to DSC. ASSD is non-negative and lower ASSD values indicate better segmentation performance.

    In addition, we evaluate the boundary alignment between the Canny edges extracted from the input-domain 2D images and the translated 2D images using the 95th percentile Hausdorff Distance (HD95). Given edge sets A (input-domain) and B (translated), HD95 is defined as:

    \begin{equation}
        HD95(A, B) = \max \Bigl(\operatorname{P}_{95}(d(A, B)), \operatorname{P}_{95}(d(B, A))\Bigr)
    \end{equation}

    where $d(A, B) = \{\min_{b \in B}\|a - b\| \;|\; a \in A\}$ denotes the set of distances from each point in A to the nearest point in B. HD95 is non-negative, with smaller values indicating better contour alignment.
    
    As there are no paired images, we also calculate the foreground\footnote{Foreground refers to pixels containing the object of interest. In this paper, we use masks from CTs to extract objects.} FID \cite{fid} and KID \cite{kid} between the translated image and output domain image distributions for reference. We do this to measure the feature alignment of the foreground object between input and output domains, free of noise from the surrounding background areas which are less important for the segmentation tasks of interest.
    
    \begin{table*}[!t]
\centering
\resizebox{1\textwidth}{!}{%
\begin{tabular}{l||cc|cc|c||cc|cc|c||cc|cc|c}
\toprule
\multicolumn{1}{c}{} & \multicolumn{5}{c}{\textbf{Lumbar}} & \multicolumn{5}{c}{\textbf{Lumbar-SPIDER}}  & \multicolumn{5}{c}{\textbf{Hip \& Thigh}} \\
\cmidrule(lr){2-6} \cmidrule(lr){7-11}\cmidrule(lr){12-16}
\multicolumn{1}{c}{} & \multicolumn{2}{c}{UNet} & \multicolumn{2}{c}{SwinUNet} & \multicolumn{1}{c}{} & \multicolumn{2}{c}{UNet} & \multicolumn{2}{c}{SwinUNet} & \multicolumn{1}{c}{} & \multicolumn{2}{c}{UNet} & \multicolumn{2}{c}{SwinUNet} & \multicolumn{1}{c}{} \\
\cmidrule(lr){2-3} \cmidrule(lr){4-5} \cmidrule(lr){7-8} \cmidrule(lr){9-10} \cmidrule(lr){12-13} \cmidrule(lr){14-15}
Method & DSC ($\uparrow$) & ASSD ($\downarrow$) & DSC ($\uparrow$) & ASSD ($\downarrow$) & Edge HD95 ($\downarrow$) & DSC ($\uparrow$) & ASSD ($\downarrow$) & DSC ($\uparrow$) & ASSD ($\downarrow$) & Edge HD95 ($\downarrow$) & DSC ($\uparrow$) & ASSD ($\downarrow$) & DSC ($\uparrow$) & ASSD ($\downarrow$) & Edge HD95 ($\downarrow$) \\
\midrule
w/o Adap. & 0.287 $\pm$ 0.034 & 6.515 $\pm$ 1.495 & 0.171 $\pm$ 0.039 & 7.386 $\pm$ 1.263 & - & 0.236 $\pm$ 0.023 & 8.275 $\pm$ 0.654 & 0.187 $\pm$ 0.022 & 8.327 $\pm$ 0.600 & - & 0.004 $\pm$ 0.002 & 45.731 $\pm$ 5.724 & 0.003 $\pm$ 0.002 & 48.624 $\pm$ 5.429 & - \\
CycleGAN & 0.484 $\pm$ 0.022 & 2.479 $\pm$ 0.160 & 0.362 $\pm$ 0.028 & 3.505 $\pm$ 0.259 & 24.673 $\pm$ 0.189 & 0.507 $\pm$ 0.015 & 3.629 $\pm$ 0.284 & 0.412 $\pm$ 0.021 & 3.701 $\pm$ 0.263 & 24.673 $\pm$ 0.189 & \underline{0.535 $\pm$ 0.038} & 9.140 $\pm$ 1.642 & \underline{0.464 $\pm$ 0.053} & \underline{9.790 $\pm$ 1.247} & 13.541 $\pm$ 0.117 \\
SynSeg-Net & 0.316 $\pm$ 0.031 & 3.013 $\pm$ 0.410 & 0.288 $\pm$ 0.040 & 3.527 $\pm$ 0.358 & 26.737 $\pm$ 0.293 & 0.364 $\pm$ 0.019 & 3.207 $\pm$ 0.197 & 0.223 $\pm$ 0.023 & 7.366 $\pm$ 0.659 & 26.737 $\pm$ 0.293 & 0.370 $\pm$ 0.064 & \underline{4.705 $\pm$ 0.456} & 0.059 $\pm$ 0.014 & 12.871 $\pm$ 1.469 & 20.123 $\pm$ 0.262 \\
CyCADA & 0.331 $\pm$ 0.024 & 5.942 $\pm$ 1.219 & 0.319 $\pm$ 0.024 & 3.691 $\pm$ 0.260 & 30.109 $\pm$ 0.262 & 0.364 $\pm$ 0.016 & 4.389 $\pm$ 0.256 & 0.260 $\pm$ 0.011 & 4.725 $\pm$ 0.186 & 30.109 $\pm$ 0.262 & 0.349 $\pm$ 0.039 & 11.247 $\pm$ 1.472 & 0.155 $\pm$ 0.033 & 13.002 $\pm$ 1.684 & \underline{12.544 $\pm$ 0.105}\\
MUNIT & 0.407 $\pm$ 0.013 & 3.803 $\pm$ 0.223 & 0.433 $\pm$ 0.016 & 3.212 $\pm$ 0.213 & 44.285 $\pm$ 0.648 & 0.380 $\pm$ 0.013 & 4.309 $\pm$ 0.290 & 0.358 $\pm$ 0.014 & 3.545 ± 0.174 & 44.285 $\pm$ 0.648 & 0.128 $\pm$ 0.026 & 16.228 $\pm$ 3.226 & 0.090 $\pm$ 0.023 & 18.925 $\pm$ 3.179 & 22.358 $\pm$ 0.205 \\
CUT & 0.392 $\pm$ 0.020 & 4.670 $\pm$ 0.745 & 0.288 $\pm$ 0.029 & 5.259 $\pm$ 0.371 & 32.665 $\pm$ 0.702 & 0.368 $\pm$ 0.020 & 5.781 $\pm$ 0.427 & 0.292 $\pm$ 0.022 & 6.751 $\pm$ 0.530 & 32.665 $\pm$ 0.702 & 0.311 $\pm$ 0.052 & 19.254 $\pm$ 3.817 & 0.211 $\pm$ 0.030 & 20.564 $\pm$ 4.384 & 27.118 $\pm$ 0.261 \\
GcGAN & \underline{0.554 $\pm$ 0.020} & \underline{1.753 ± 0.087} & 0.433 $\pm$ 0.030 & \underline{2.940 $\pm$ 0.372} & \underline{11.683 $\pm$ 0.216} & \underline{0.580 $\pm$ 0.010} & \underline{2.202 $\pm$ 0.096} & \underline{0.513 $\pm$ 0.013} & \underline{2.904 $\pm$ 0.157} & \underline{11.683 $\pm$ 0.216} & 0.414 $\pm$ 0.048 & 9.275 $\pm$ 2.035 & 0.320 $\pm$ 0.043 & 13.650 $\pm$ 2.459 & 25.998 $\pm$ 0.271 \\
MaskGAN & 0.428 $\pm$ 0.026 & 3.192 $\pm$ 0.251 & 0.322 $\pm$ 0.039 & 4.692 $\pm$ 0.917 & 16.961 $\pm$ 0.213 & 0.458 $\pm$ 0.017 & 3.729 $\pm$ 0.253 & 0.385 $\pm$ 0.023 & 5.355 $\pm$ 0.438 & 16.961 $\pm$ 0.213 & 0.289 $\pm$ 0.048 & 16.229 $\pm$ 3.576 & 0.292 $\pm$ 0.032 & 17.590 $\pm$ 3.245 & 30.838 $\pm$ 0.276 \\
FGDM & 0.455 $\pm$ 0.022 & 4.658 $\pm$ 0.727 & 0.390 $\pm$ 0.022 & 5.589 $\pm$ 0.783 & 15.701 $\pm$ 0.167 & 0.411 $\pm$ 0.021 & 5.077 $\pm$ 0.441 & 0.333 $\pm$ 0.025 & 6.348 $\pm$ 0.510 & 15.701 $\pm$ 0.167 & 0.074 $\pm$ 0.019 & 31.927 $\pm$ 5.861 & 0.070 $\pm$ 0.020 & 29.386 $\pm$ 5.170 & 38.282 $\pm$ 0.258 \\
UNSB & 0.465 $\pm$ 0.028 & 3.111 $\pm$ 0.263 & \underline{0.456 $\pm$ 0.016} & 2.954 $\pm$ 0.206 & 29.499 $\pm$ 0.645 & 0.488 $\pm$ 0.014 & 3.984 $\pm$ 0.437 & 0.446 $\pm$ 0.017 & 3.070 $\pm$ 0.132 & 29.499 $\pm$ 0.645 & 0.247 $\pm$ 0.035 & 13.426 $\pm$ 2.399 & 0.181 $\pm$ 0.041 & 17.650 $\pm$ 3.717 & 19.101 $\pm$ 0.187 \\
\textbf{Ours} & \textbf{0.705 $\pm$ 0.019} & \textbf{1.677 $\pm$ 0.507} & \textbf{0.669 $\pm$ 0.019} & \textbf{1.570 $\pm$ 0.217} & \textbf{3.396 $\pm$ 0.020} & \textbf{0.655 $\pm$ 0.011} & \textbf{1.955 $\pm$ 0.153} & \textbf{0.603 $\pm$ 0.016} & \textbf{2.226 $\pm$ 0.191} & \textbf{3.396 $\pm$ 0.020} & \textbf{0.769 $\pm$ 0.036} & \textbf{2.625 $\pm$ 0.533} & \textbf{0.684 $\pm$ 0.060} & \textbf{4.258 $\pm$ 0.977} & \textbf{3.578 $\pm$ 0.078} \\
\midrule
UB$^\dag$ & 0.748 $\pm$ 0.021 & 1.296 $\pm$ 0.182 & 0.740 $\pm$ 0.021 & 1.315 $\pm$ 0.162 & - & 0.764 $\pm$ 0.005 & 1.325 $\pm$ 0.040 & 0.765 $\pm$ 0.004 & 1.421 $\pm$ 0.057 & - & 0.857 $\pm$ 0.020 & 1.678 $\pm$ 0.283 & 0.786 $\pm$ 0.042 & 2.992 $\pm$ 0.569 & - \\
\bottomrule
\end{tabular}%
}
\caption{Quantitative comparison (DSC, ASSD and Edge HD95) of ContourDiff to other image translation methods in terms of segmentation model performance on held-out output domain images. (L: Lumbar dataset, L-SPIDER: SPIDER Lumbar dataset, H \& T: Hip \& Thigh dataset). ``w/o Adap.'' is the baseline referring to the model trained on CTs without any adaptation and tested on MRIs directly. UB$^\dag$ represents the upper bound model trained on real annotated output domain images. Best in bold, runner-up underlined. Standard error of the mean (SEM) is reported with each result.
}
\label{tab:result_dsc_assd}
\end{table*}
    \begin{table*}[h!]
\centering
\resizebox{1\textwidth}{!}{%
\begin{tabular}{c|c|c|c|c|c|c|c|c|c|c}
\toprule
\multicolumn{1}{c}{} & \multicolumn{9}{c}{\textbf{Lumbar Spine (L) - Foreground}} \\
\cmidrule(lr){2-11}
\textbf{Metric} & CycleGAN & SynSeg-Net & CyCADA & MUNIT & CUT & GcGAN & MaskGAN & FGDM & UNSB & \textbf{Ours}  \\
\midrule
FID ($\downarrow$) & 132.16 & 137.63 & \underline{127.54} & 372.67 & 150.10 & 138.60 & 128.17 & 158.69 & 137.42 & \textbf{126.35} \\
KID ($\downarrow$) & 0.047 & 0.054 & 0.045 & 0.343 & 0.058 & 0.050 & \textbf{0.039} & 0.071 & 0.051 & \underline{0.042} \\
\midrule
\multicolumn{1}{c}{} & \multicolumn{9}{c}{\textbf{Hip \& Thigh (H \& T) - Foreground}} \\
\cmidrule(lr){2-11}
\textbf{Metric} & CycleGAN & SynSeg-Net & CyCADA & MUNIT  & CUT & GcGAN & MaskGAN & FGDM & UNSB & \textbf{Ours}  \\
\midrule
FID ($\downarrow$) & 183.18 & 192.32 & 184.11 & 193.12 & 193.63 & \underline{163.61} & 175.28 & 251.85 & 167.88 & \textbf{133.74} \\
KID ($\downarrow$) & 0.163 & 0.169 & 0.159 & 0.174 & 0.178 & 0.144 & 0.152 & 0.257 & \underline{0.142} & \textbf{0.093} \\

\bottomrule
\end{tabular}%
}
\caption{Quantitative comparison of foreground FID and KID between translated images and output domain images. Best in bold, runner-up underlined. (L-SPIDER is excluded as it is only used for testing and not for training.)}
\label{tab:result_fid_kid}
\end{table*}

    \subsection{Implementation Details}
    \subsubsection{Model Architecture}
    For the image translation model, we adopt the UNet architecture \cite{unet} for the denoising model $\epsilon_{\theta}$ with a three-channel input (grayscale image, its contour and spatial information from a grayscale neighbor slice). The encoder comprises six down sampling stages, each consisting of two ResNet blocks. The number of output feature channels at each stage is (128, 128, 256, 256, 512, 512). To capture long-range dependencies, we integrate spatial self-attention at the fifth stage. The decoder symmetrically upsamples through six stages, fusing encoder features via skip connections, and produces a single-channel translated image at the same resolution as the input. For canny edge extraction, we inspected a representative subset of images and selected low/high thresholds of 30/50 for CT and 50/100 for MRI to optimally capture anatomical contours while suppressing spurious edges. We primarily conduct experiments with $P_{adj}=0.2$ and evaluate other values in the ablation studies. All experiments are running on a NVIDIA RTX A6000 GPU.
    
    \subsubsection{Model Training}
    The training settings for the diffusion model follow the same as that in \cite{konz2024anatomically}. We use the DDIM algorithm \cite{song2021ddim} for sampling, with 50 steps. For the segmentation models, we use the convolution-based UNet \cite{unet} and transformer-based SwinUNet \cite{cao2022swin}. All images are resized to $256 \times 256$ and normalized to 8-bit $[0, 255]$, following common preprocessing practices in medical imaging analysis \cite{mazurowski2023segment,ma2024segment,lyu2024superpixel,konz2024anatomically}.  The training of competing methods mostly follows the default settings from each official GitHub \footnote{Training time: 200 epochs (CycleGAN, SynSeg-Net, CyCADA, GcGAN, MaskGAN); 400 epochs (CUT); 1M iterations (MUNIT); 60/280 epochs (UNSB on L/H \& T).}. We set the $\lambda_{idt} = 0.5$ to include identity loss if provided.  We train the segmentation model with a cosine learning rate scheduler up to 100 epochs with the initial learning rate of $1 \times 10^{-3}$.

    \begin{figure*}[!ht]
        \centering
        \includegraphics[width=1\textwidth]{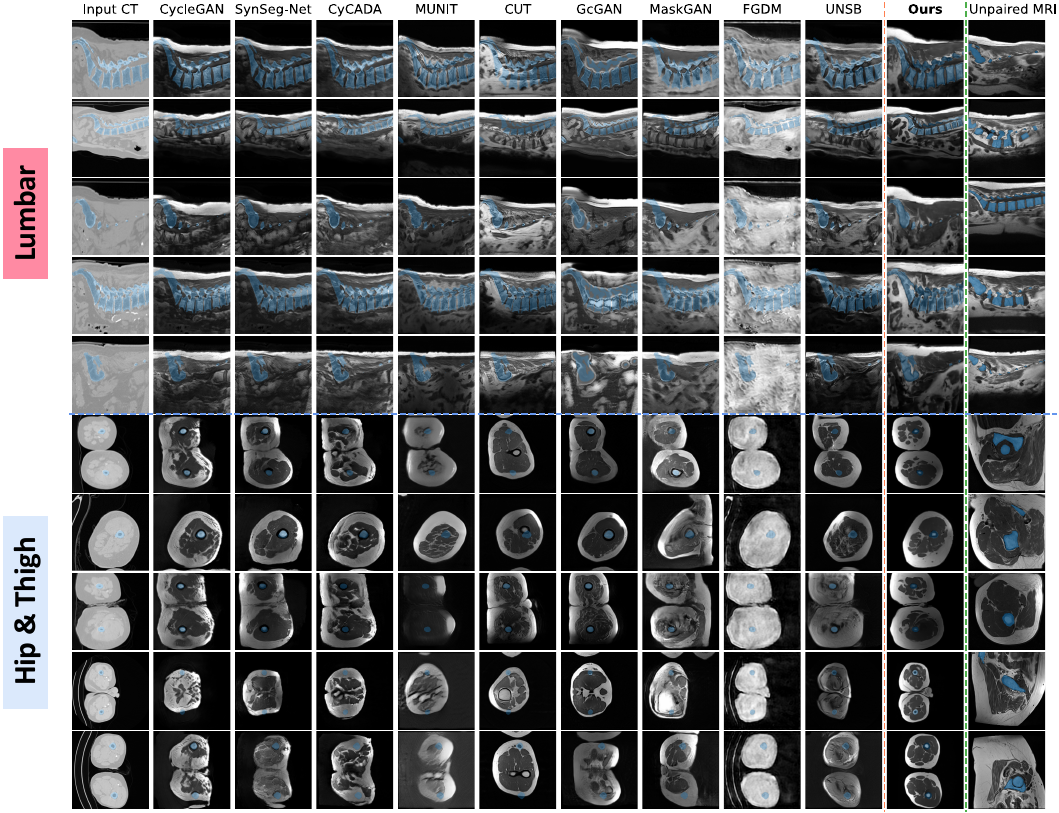}
        \caption{\textbf{Qualitative comparison of ContourDiff and baseline methods.} ContourDiff appears to best maintain anatomical consistency during translation for both Lumbar and Hip $\&$ Thigh areas. The input-domain segmentation masks are depicted in blue to visualize the alignment. Unpaired MRIs are included as target-domain examples for reference only. Note: they are no used as ground truth for the translation.}
        \label{fig:translation}
    \end{figure*}
    
    \subsection{Comparison with Other Methods}
    We compare our framework to other translation/adaptation methods, including CycleGAN \cite{cyclegan}, SynSeg-Net \cite{synsegnet}, CyCADA \cite{cycada}, MUNIT \cite{munit}, CUT \cite{cut}, GcGAN \cite{gcgan}, MaskGAN \cite{maskgan}, FGDM \cite{li2023zero} and UNSB \cite{unsb}, via the performance of output domain-trained downstream task segmentation models on translated images. Several of these methods (e.g., \cite{cyclegan,synsegnet,munit,cut,gcgan,maskgan,li2023zero,unsb}) translate the images solely at the image level, while CyCADA also aligns latent features from the downstream task encoder. MaskGAN uses the extracted coarse masks to better preserve object structures throughout translation. In addition to GAN-based models, FGDM utilizes both low and high frequency information as diffusion conditions for translation, and UNSB integrates diffusion models with Schr\"{o}dinger Bridge theory to enable probabilistically consistent translation for unpaired data. For CyCADA, we used the same segmentation architecture as other methods but without the skip connection to enable feature alignment. For each competing method, we evaluated multiple intermediate results from the translation \footnote{For SynSeg-Net and CyCADA, we evaluate the segmentation model every 20 epochs. For other methods, as we need to train the segmentation model separately, we evaluate at 10\%, 30\%, 50\%, 75\% and 100\% of the training time.} and report the best performance.
    
    \subsection{Results}
    \subsubsection{Quantitative Results}
    The segmentation model results are shown in Table \ref{tab:result_dsc_assd}. Overall, across all three test sets, our method consistently outperforms previous image translation methods by a significant margin on both DSC and ASSD, using either a UNet or a SwinUNet. Specifically, as for UNet, our method achieves DSC improvements of at least $15.1\%$, $7.5\%$ and $23.4\%$ on the L, L-SPIDER and H\&T datasets, respectively. When using SwinUNet, the DSC improvements of segmentation model are at least $21.3\%$, $9\%$ and $22\%$ on the same three datasets. These results demonstrate the superior translation performance of our method in developing cross-domain segmentation models. Furthermore, our method significantly outperforms all baselines in terms of edge alignment, reducing HD95 by 8.287 and 8.966 in compared with the runner-up methods in lumbar and hip \& thigh regions, respectively, indicating better anatomical contour alignment during translation.
    
    Based on Table \ref{tab:result_fid_kid}, our method achieves the lowest FID scores: 126.35 and 133.74 for L and H \& T, respectively. For KID scores, our method outperforms others for H \& T and achieves a close second place for L (0.042), which is slightly lower than the top score of 0.039 by MaskGAN. The improvements in FID and KID demonstrate the superior foreground fidelity of the images translated by our method compared to other baselines.
    
    \subsubsection{Qualitative Results} 
    We provide example translated images in Fig. \ref{fig:translation}. These datasets form a challenging task due to (1) the noticeable shift in image features between the input and output domains and (2) the high anatomical variability between different scans. Moreover, we see that adversarially-trained models (e.g., CycleGAN) have trouble with the consistent structural shift (i.e., large structural bias) between the input and output domains, i.e., when one domain is absent of certain features seen in the other. As shown in Fig. \ref{fig:teaser}, this is particularly evident in our H\&T dataset, where MRIs are dominant by a single leg, and CTs often contain two legs. Such a bias may lead the adversarial mechanism to over-emphasize these features and, therefore, tend to translate CTs of two legs into MRIs depicting only one leg (see Fig. \ref{fig:translation}). For the lumbar spine from the sagittal view, MRIs often start from the lowest thoracic spine and end at the sacrum. On the other hand, CTs often include the upper leg and sometimes the abdominal body (see Fig. \ref{fig:translation}). 
    
    Fig. \ref{fig:translation} shows that our model explicitly enforces anatomical consistency through translation despite these domain feature differences through its contour guidance, generating MRIs that strictly follow input CT images, resulting in better mask alignment and better segmentation model performance. Notably, the translated outputs from ContourDiff also preseve clinically relevant anatomical details, such as clear boundaries between fat and muscle, as well as other major structures.
    
    Based on Table \ref{tab:result_dsc_assd}, Table \ref{tab:result_fid_kid} and Fig. \ref{fig:translation}, ContourDiff best maintain anatomical fidelity and consistency compared to other models, both quantitatively and qualitatively.
    
    \subsection{Ablation Studies}
    We will now conduct ablation studies to validate the effectiveness of key design choices for ContourDiff, studying how contours are used for guidance, and the general effectiveness of SCGD and its dependence on $P_{adj}$.
    
    \subsubsection{Effectiveness of Adding Contours}
\begin{figure}[!ht]
        \centering
        \includegraphics[width=1\columnwidth]{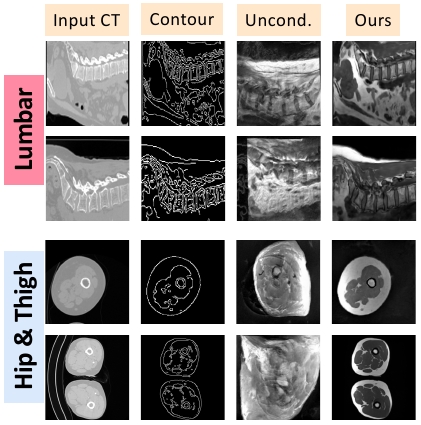}
        \caption{\textbf{Qualitative comparison between unconditional DDPM and ContourDiff.} Unconditional DDPM seems to hardly follow input-domain anatomical structures during translation.}
        \label{fig:ablation_unconditional}
    \end{figure}

    We verify the effectiveness of introducing contours to each denoising step during training by conditionally training on an empty map (i.e., all zeros) and adding the CT contours during the translation steps. Fig. \ref{fig:ablation_unconditional} showed that the denoised model $\epsilon_\theta$ trained without contours hardly followed the introduced CTs contours (\textbf{\lq Uncond.\rq} column). Furthermore, the UNet trained on these unconditionally generated MRIs experienced a dramatic performance drop (see Table \ref{tab:result_ablation_contour}). These results demonstrate the necessity of including contours to achieve anatomical consistency during translation.
    \begin{table}[h!]
\centering
\resizebox{1\columnwidth}{!}{%
\begin{tabular}{l|cc|cc|cc}
\toprule
\multicolumn{1}{c}{} & \multicolumn{2}{c}{\textbf{L}} & \multicolumn{2}{c}{\textbf{L-SPIDER}}  & \multicolumn{2}{c}{\textbf{H \& T}} \\
\cmidrule(lr){2-3} \cmidrule(lr){4-5}\cmidrule(lr){6-7}
Method & DSC ($\uparrow$) & ASSD ($\downarrow$) & DSC ($\uparrow$) & ASSD ($\downarrow$) & DSC ($\uparrow$) & ASSD ($\downarrow$) \\
\midrule
Unconditional & 0.264 $\pm$ 0.044 & 5.148 $\pm$ 1.337 & 0.202 $\pm$ 0.023 & 6.479 $\pm$ 0.595 & 0.298 $\pm$ 0.032 & 19.352 $\pm$ 4.153 \\
Ours & \textbf{0.705 $\pm$ 0.019} & \textbf{1.677 $\pm$ 0.507} & \textbf{0.655 $\pm$ 0.011} & \textbf{1.955 $\pm$ 0.153} & \textbf{0.769 $\pm$ 0.036} & \textbf{2.625 $\pm$ 0.533} \\
\bottomrule
\end{tabular}%
}
\caption{Quantitative comparison (DSC and ASSD) of ContourDiff and unconditional DDPM. Best in bold.}
\label{tab:result_ablation_contour}
\end{table}

    \begin{figure}[!ht]
        \centering
        \includegraphics[width=1\columnwidth]{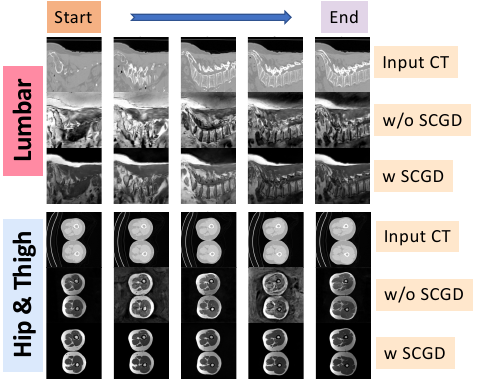}
        \caption{\textbf{Qualitative results of translation with SCGD.} SCGD can better preserve the consistency between translated slices within volume.}
        \label{fig:ablation_scgd}
    \end{figure}

    \begin{figure}[!ht]
        \centering
        \includegraphics[width=1\columnwidth]{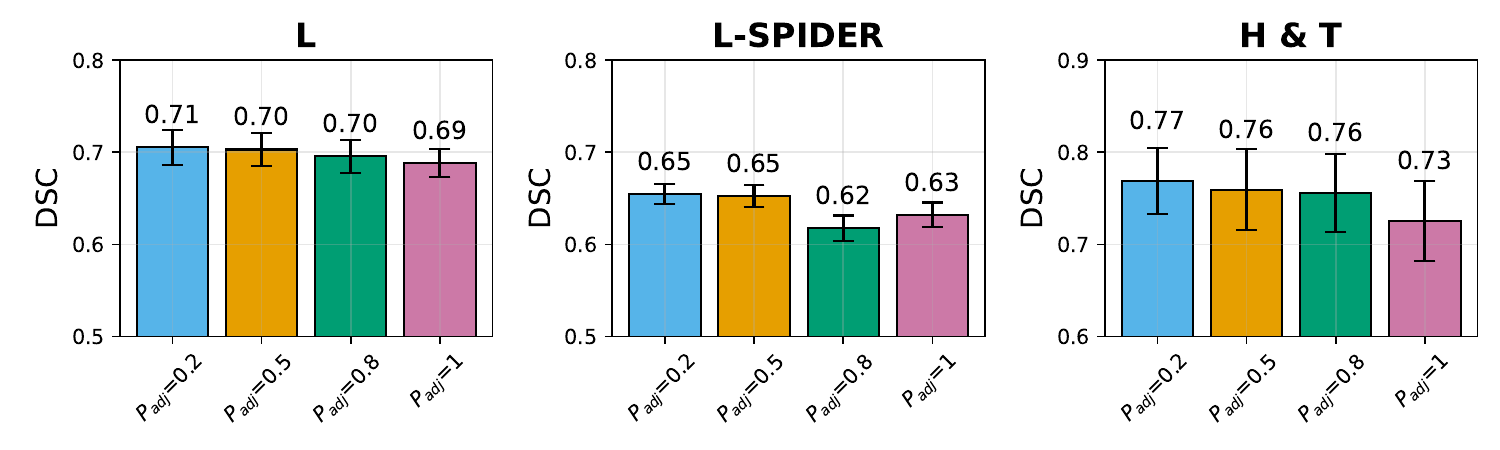}
        \caption{\textbf{Quantitative comparison of different choices on $P_{adj}$.} Smaller $P_{adj}$ seems to result in better translation performance.}
        \label{fig:ablation_bar_plot}
    \end{figure}

    \begin{figure*}[!t]
        \centering
        \includegraphics[width=2\columnwidth]{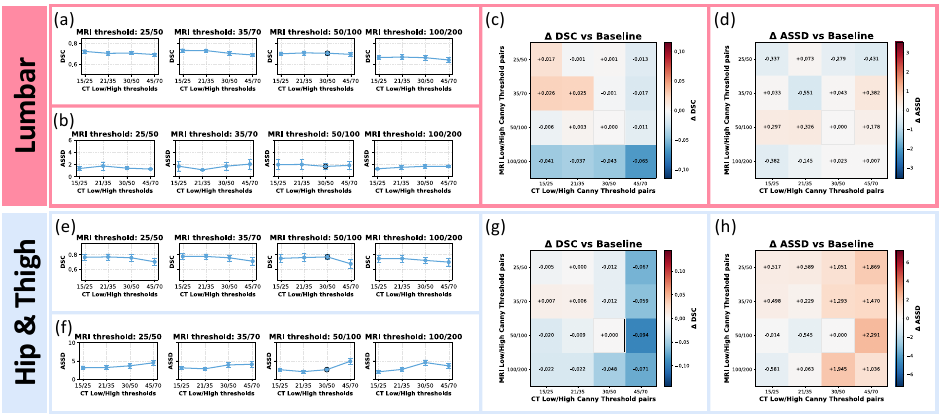}
        \caption{\textbf{Model performance under different Canny threshold pairs.} (a), (e) and (b), (f) show DSC and ASSD with standard error of the mean (SEM) across threshold pairs. (c), (g) and (d), (h) present heatmaps of the corresponding DSC and ASSD deltas relative to the baseline setting.}
        \label{fig:robustness_vs_threshold}
    \end{figure*}

    \subsubsection{Effectiveness of SCGD}
    To assess the impact of the proposed SCGD, we also generate translated images without guidance from the adjacent slice. Similarly, we then train a UNet and report the segmentation performance with the standard error of the mean (SEM). As shown in Table \ref{tab:result_ablation_scgd} and Fig. \ref{fig:ablation_scgd}, ContourDiff with SCGD achieves a superior performance compared to that without SCGD (i.e., without guidance from adjacent slices). Moreover, according to Table \ref{tab:result_dsc_assd} and \ref{tab:result_ablation_scgd}, ContourDiff without SCGD still outperforms baseline methods by a significant margin, further demonstrating its promise for preserving anatomical structures with 2D data.
    \begin{table}[!h]
\centering
\resizebox{1\columnwidth}{!}{%
\begin{tabular}{l|cc|cc|cc}
\toprule
\multicolumn{1}{c}{} & \multicolumn{2}{c}{\textbf{L}} & \multicolumn{2}{c}{\textbf{L-SPIDER}}  & \multicolumn{2}{c}{\textbf{H \& T}} \\
\cmidrule(lr){2-3} \cmidrule(lr){4-5}\cmidrule(lr){6-7}
Method & DSC ($\uparrow$) & ASSD ($\downarrow$) & DSC ($\uparrow$) & ASSD ($\downarrow$) & DSC ($\uparrow$) & ASSD ($\downarrow$) \\
\midrule
w/o SCGD & 0.653 $\pm$ 0.025 & 2.425 $\pm$ 0.758 & 0.583 $\pm$ 0.021 & 2.315 $\pm$ 0.221 & 0.706 $\pm$ 0.044 & 3.489 $\pm$ 0.540 \\
Ours & \textbf{0.705 $\pm$ 0.019} & \textbf{1.677 $\pm$ 0.507} & \textbf{0.655 $\pm$ 0.011} & \textbf{1.955 $\pm$ 0.153} & \textbf{0.769 $\pm$ 0.036} & \textbf{2.625 $\pm$ 0.533} \\
\bottomrule
\end{tabular}%
}
\caption{Quantitative comparison (DSC and ASSD) of ContourDiff with SCGD and without SCGD. Best in bold.}
\label{tab:result_ablation_scgd}
\end{table}
    
    \subsubsection{Different Choices on $P_{adj}$}
    We now investigate how varying $P_{adj}$ affects translation quality by training and evaluating the UNet architecture for each setting. As shown in Table \ref{tab:result_ablation_prob} and Fig. \ref{fig:ablation_bar_plot}, lower $P_{adj}$ values seem to result in better translation performance. This outcome is expected because a smaller $P_{adj}$ forces the model to rely more heavily on anatomical contour constraints. By contrast, a larger $P_{adj}$ may introduce conflicting contextual information from adjacent slices and may encourage the network to learn direct mappings from adjacent slices instead of preserving true anatomical fidelity from contours. Thus, in practice, we recommend choosing $P_{adj}$ at or below 0.5.
    \begin{table}[h!]
\centering
\resizebox{1\columnwidth}{!}{%
\begin{tabular}{l|cc|cc|cc}
\toprule
\multicolumn{1}{c}{} & \multicolumn{2}{c}{\textbf{L}} & \multicolumn{2}{c}{\textbf{L-SPIDER}}  & \multicolumn{2}{c}{\textbf{H \& T}} \\
\cmidrule(lr){2-3} \cmidrule(lr){4-5}\cmidrule(lr){6-7}
$P_{adj}$ & DSC ($\uparrow$) & ASSD ($\downarrow$) & DSC ($\uparrow$) & ASSD ($\downarrow$) & DSC ($\uparrow$) & ASSD ($\downarrow$) \\
\midrule
1 & 0.688 $\pm$ 0.016 & 1.555 $\pm$ 0.284 & 0.632 $\pm$ 0.013 & 2.279 $\pm$ 0.203 & 0.726 $\pm$ 0.044 & 4.480 $\pm$ 0.732 \\
0.8 & 0.696 $\pm$ 0.018 & \textbf{1.194 $\pm$ 0.099} & 0.618 $\pm$ 0.014 & \textbf{1.902 $\pm$ 0.127} & 0.756 $\pm$ 0.043 & 3.285 $\pm$ 0.637 \\
0.5 & \underline{0.703 $\pm$ 0.018} & \underline{1.393 $\pm$ 0.277} & \underline{0.653 $\pm$ 0.012} & \underline{1.949 $\pm$ 0.147} & \underline{0.760 $\pm$ 0.044} & \underline{2.639 $\pm$ 0.670} \\
Ours & \textbf{0.705 $\pm$ 0.019} & 1.677 $\pm$ 0.507 & \textbf{0.655 $\pm$ 0.011} & 1.955 $\pm$ 0.153 & \textbf{0.769 $\pm$ 0.036} & \textbf{2.625 $\pm$ 0.533} \\
\bottomrule
\end{tabular}%
}
\caption{Quantitative results (DSC and ASSD) of ablation study in terms of segmentation model performance to explore different $P_{adj}$ values. Best in bold, runner-up underlined.}
\label{tab:result_ablation_prob}
\end{table}

\subsection{Experiments on T2 MRI to T1 MRI}
\label{sec:t2tot1}
    To further verify the zero-shot capability of ContourDiff, we incorporate an additional experiment on translating T2-weighted MRI to T1-weighted MRI in the hip and thigh region. We collect 594 annotated 2D T2 MRI slices, and split into 502:92 for training and validation. We directly apply the ContourDiff model previously trained for CT to MRI translation to perform T2 to T1 translation in a zero-shot manner, followed by downstream segmentation evaluation using UNet on the same test set as used for CT to MRI tasks.

    \begin{table}[!h]
\centering
\resizebox{0.75\columnwidth}{!}{%
\begin{tabular}{l|ccc}
\toprule
\multicolumn{1}{c}{} & \multicolumn{3}{c}{\textbf{H \& T (T2 MRI$\rightarrow$T1 MRI)}} \\
\cmidrule(lr){2-4}
Method & DSC ($\uparrow$) & ASSD ($\downarrow$) & Edge HD95 ($\downarrow$) \\
\midrule
w/o Adap. & 0.647 $\pm$ 0.062 & 7.810 $\pm$ 2.313 & - \\
CycleGAN & 0.714 $\pm$ 0.047 & 4.669 $\pm$ 0.811 & 25.418 $\pm$ 0.708 \\
UNSB & \underline{0.724 $\pm$ 0.048} & \underline{4.366 $\pm$ 0.938} & \underline{18.649 $\pm$ 0.706} \\
\textbf{Ours} & \textbf{0.778 $\pm$ 0.040} & \textbf{2.927 $\pm$ 0.472} & \textbf{5.699 $\pm$ 0.154} \\
\midrule
UB$^\dag$ & 0.857 $\pm$ 0.020 & 1.678 $\pm$ 0.283 & - \\
\bottomrule
\end{tabular}%
}
\caption{Quantitative comparison (DSC, ASSD and Edge HD95) of ContourDiff on translating T2-weighted MRI to T1-weighted MRI. Best in bold, runner-up underlined.}
\label{tab:result_leg_t2}
\end{table}

    For comparison, we include representative GAN-based (CycleGAN) and diffusion-based (UNSB) baselines. As shown in Table \ref{tab:result_leg_t2}, ContourDiff model not only achieves superior segmentation performance on T2 to T1 translation but also best aligns edges, all without any additional training, demonstrating its zero-shot capability.

\subsection{Model Robustness Evaluation}
    We evaluate the robustness of ContourDiff in terms of different Canny thresholds, image qualities, and image contrasts.

    \subsubsection{Model Robustness under Different Canny Thresholds}
    We select three threshold pairs around the current settings for both CTs (30/50) and MRIs (50/100), and evaluate segmentation performance using UNet on both L and H\&T datasets. Specifically, we consider low/high threshold pairs of 15/25, 21/35, and 45/70 for CT, and 25/50, 35/70, and 100/200 for MRI.

    As shown in Figure \ref{fig:robustness_vs_threshold}, ContourDiff mostly exhibits robust performance across different Canny threshold pairs. Lower thresholds, which produce more detailed contour maps, seem to generally maintain or even slightly improve performance. In contrast, higher thresholds reduce the number of detected contours and occasionally lead to small performance drops. This is reasonable as sparse edge information may weaken anatomical guidance. In particular, even with minor performance degradation, ContourDiff still significantly outperforms all existing baselines (compared to Table \ref{tab:result_dsc_assd}).

    \subsubsection{Model Robustness under Different Image Qualities}
    We simulate variations in image quality by adding Gaussian noise to the original images at signal-to-noise ratio (SNR) levels. Specifically, given a desired SNR in decibels ($SNR_{dB}$), the noise power is:
    \begin{equation}
        P_{noise} = \frac{P_{signal}}{10^{\frac{SNR_{dB}}{10}}}
    \end{equation}

    where $P_{signal}$ is the mean sqaured intensity of the original image and $P_{noise}$ is the variance of added noise. We add noise at three SNR levels (30 dB, 25 dB, and 15 dB) to the input CT images and evaluate segmentation performance using UNet for both L and H\&T datasets.

    \begin{figure}[!ht]
        \centering
        \includegraphics[width=1\columnwidth]{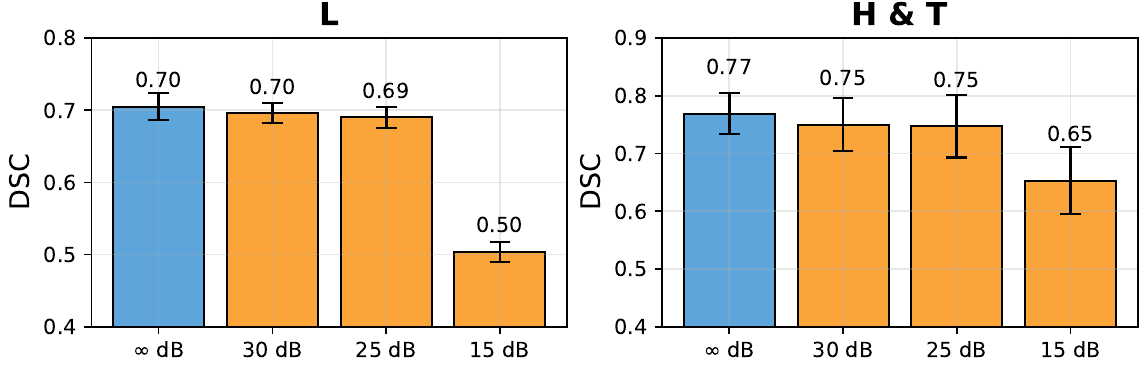}
        \caption{\textbf{Model performance under different image qualities.} $\infty$ dB represents the original images without any added noise.}
        \label{fig:robustness_vs_snr}
    \end{figure}

    \begin{figure}[!ht]
        \centering
        \includegraphics[width=1\columnwidth]{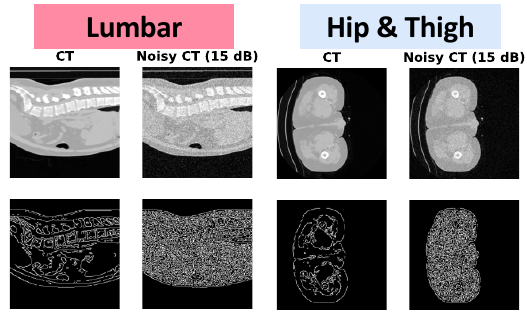}
        \caption{\textbf{Image examples and extracted contours under severe noise.} Contours extracted from heavily degraded images tend to lose anatomical details.}
        \label{fig:snr_example}
    \end{figure}

    Based on Figure \ref{fig:robustness_vs_snr}, the ContourDiff model remains robust under mild-to-moderate noise levels (e.g., 30 dB and 25 dB), with at most a 0.02 DSC drop. Moreover, the performance degrades under more severe noise (e.g., 15 dB), as the extracted contours under such conditions contain limited anatomical details (Figure \ref{fig:snr_example}).

    \subsubsection{Model Robustness under Different Image Contrasts}
    We adjust contrasts of the CT images and MRI images to assess model robustness. A linear contrast transformation is applied to the original images as follows:

    \begin{equation}
        I^{'} = c + k \times (I - c)
    \end{equation}

    where $I$ is the original pixel intensity, $c$ is the mean intensity of each image, and $k$ is the contrast factor. We apply the contrast transformation to only one modality at a time while keeping the other modality at its original contrast. We evaluate segmentation performance on both L and H\&T using UNet under two contrast settings: $k=0.8$ (reduced contrast) and $k=1.2$ (enhanced contrast).

    \begin{figure}[!ht]
        \centering
        \includegraphics[width=1\columnwidth]{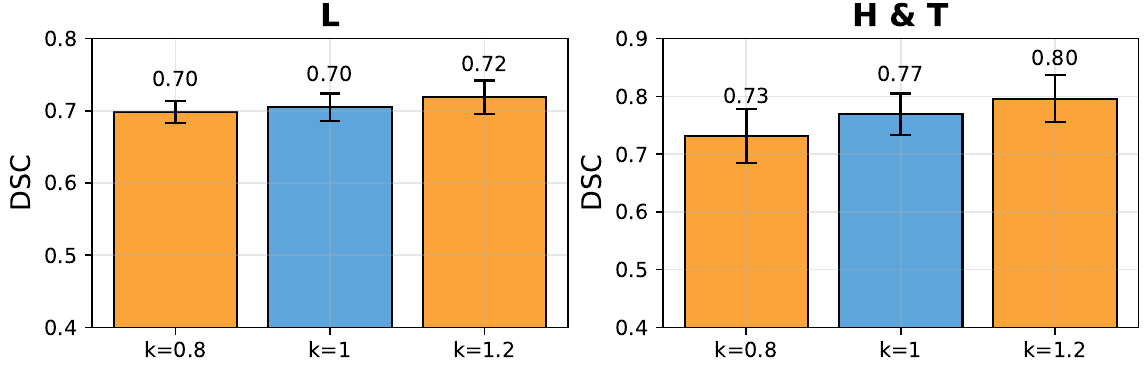}
        \caption{\textbf{Model performance under different CT contrast levels (with original MRI).} $k=1$ represents the original images without any contrast changes.}
        \label{fig:robustness_vs_contrast}
    \end{figure}

    \begin{figure}[!ht]
        \centering
        \includegraphics[width=1\columnwidth]{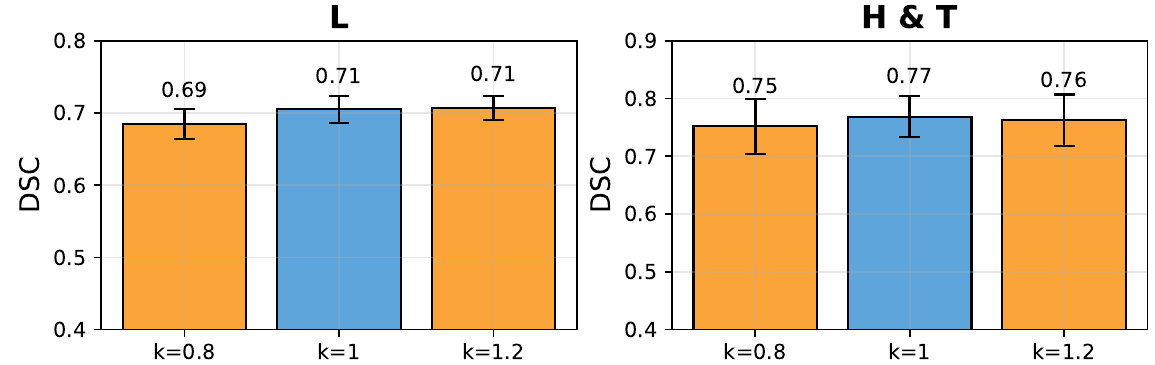}
        \caption{\textbf{Model performance under different MRI contrast levels (with original CT).} $k=1$ represents the original images without any contrast changes.}
        \label{fig:robustness_vs_mri_contrast}
    \end{figure}

    As shown in Figure \ref{fig:robustness_vs_contrast} and \ref{fig:robustness_vs_mri_contrast}, the performance of ContourDiff remains largely stable, with slight improvements with enhanced contrast and minor degradations with reduced contrast. This is expected, as higher contrast produces clearer extracted contours and preserves more anatomical details. Therefore, in practice, enhancing contrast for both input-domain and output-domain images prior to applying the method may lead to better performance.

    \subsubsection{Sampling Stability Assessment}
    \begin{figure}[!ht]
        \centering
        \includegraphics[width=1\columnwidth]{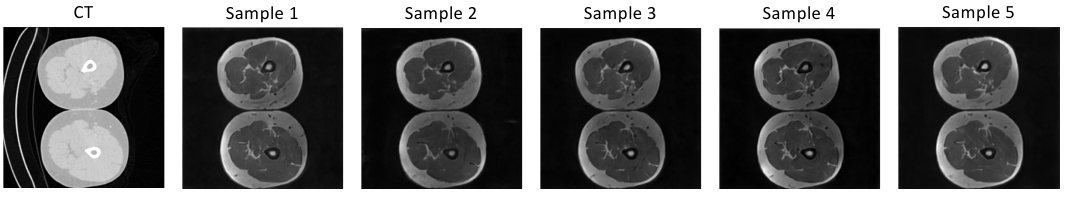}
        \caption{\textbf{Resamples visualization.} Multiple translated samples conditioned on the same input-domain information.}
        \label{fig:resample}
    \end{figure}
    
    We further evaluate the sampling stability of our model. First, we manually inspect 10 directly translated volume without any slice selection and find that, on average, 16.92\% of slices within each volume exhibit overly bright backgrounds. Next, we quantitatively assess sampling consistency across stochastic runs by computing the mean pixel-wise variance across multiple translated samples per slice. Specifically, we calculate the variance over 100 randomly selected CT contours, each with 10 generated samples, and the output mean variance is 0.007. As shown in Figure \ref{fig:resample}, translated samples from the same input-domain conditions under different seeds remain anatomically stable. The modest variability suggests some residual sampling instability. Therefore, to further stress test the initial slice selection process, we performed experiments by varying (1) the number of candidates for initial slice selection, and (2) the selection of suboptimal candidates for initial slice.

    \begin{figure}[!ht]
        \centering
        \includegraphics[width=1\columnwidth]{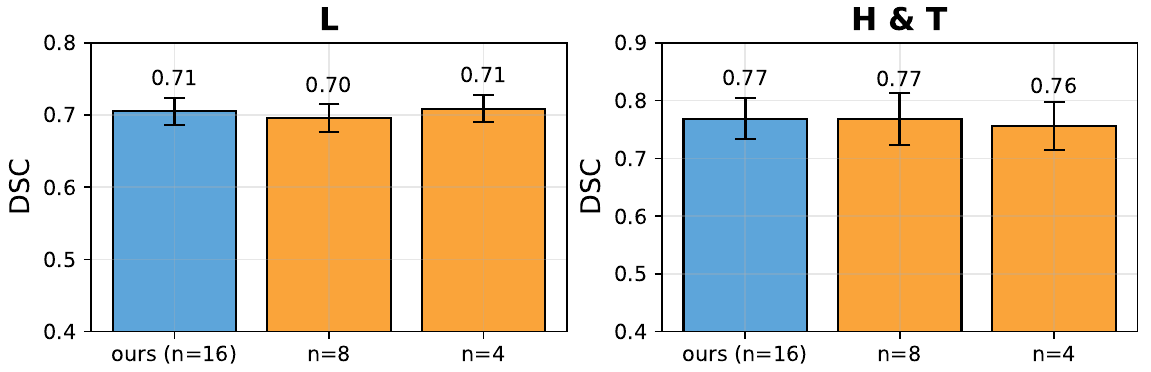}
        \caption{\textbf{Model performance with varying number of candidate slices for initial slice selection.} $k=16$ represents the current setting.}
        \label{fig:robustness_vs_n}
    \end{figure}

    \begin{figure}[!ht]
        \centering
        \includegraphics[width=1\columnwidth]{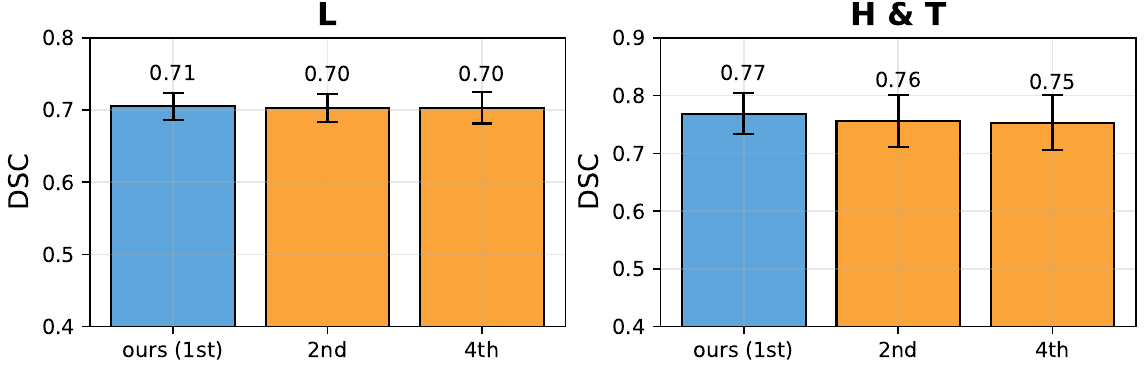}
        \caption{\textbf{Model performance when selecting different ranks of initial slice candidates.} $1st$ represents the current setting.}
        \label{fig:robustness_vs_sub}
    \end{figure}

    As shown in Figure \ref{fig:robustness_vs_n} and \ref{fig:robustness_vs_sub}, although sampling can introduce some instability, ContourDiff demonstrates strong robustness when fewer candidate slices are available or when suboptimal candidates are selected, with a DSC reduction of at most 0.02.

    \subsection{Experiments on Liver Translation}
    We perform CT to MRI liver translation using the public AMOS dataset \cite{ji2022amos} to further demonstrate the capability of ContourDiff on soft-tissue structures. Following previous settings, we collect 1,077 2D axial MRI slices for training ContourDiff model. For the downstream liver segmentation task, we randomly split 2,126 2D axial CT slices by volume into 1,625 for training and 501 for validation, and hold out 695 2D axial MRI slices as the test set.

    \begin{table}[!h]
\centering
\resizebox{0.75\columnwidth}{!}{%
\begin{tabular}{l|ccc}
\toprule
\multicolumn{1}{c}{} & \multicolumn{2}{c}{\textbf{Liver}} \\
\cmidrule(lr){2-4}
Method & DSC ($\uparrow$) & ASSD ($\downarrow$) & Edge HD95 ($\downarrow$) \\
\midrule
w/o Adap. & 0.299 $\pm$ 0.031 & 8.545 $\pm$ 0.641 & - \\
CycleGAN & 0.848 $\pm$ 0.027 & 2.419 $\pm$ 0.380 & 20.787 $\pm$ 0.281 \\
UNSB & \underline{0.872 $\pm$ 0.008} & \underline{2.377 $\pm$ 0.208} & \underline{16.962 $\pm$ 0.157} \\
\textbf{Ours} & \textbf{0.873 $\pm$ 0.012} & \textbf{2.193 $\pm$ 0.211} & \textbf{4.451 $\pm$ 0.069} \\
\midrule
UB$^\dag$ & 0.913 $\pm$ 0.015 & 1.404 $\pm$ 0.137 & - \\
\bottomrule
\end{tabular}%
}
\caption{Quantitative comparison (DSC, ASSD and Edge HD95) of ContourDiff on CT to MRI liver translation task. Best in bold, runner-up underlined.}
\label{tab:result_liver}
\end{table}

    \begin{figure}[!ht]
        \centering
        \includegraphics[width=1\columnwidth]{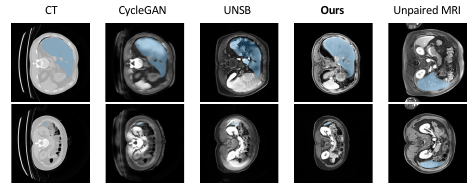}
        \caption{\textbf{Qualitative comparison of ContourDiff on liver translation.} ContourDiff seems to best keep anatomical consistency during translation for soft-tissue structures in abdominal area. The input-domain segmentation masks are presented in blue to visualize the alignment.}
        \label{fig:liver}
    \end{figure}

    According to Table \ref{tab:result_liver} and Figure \ref{fig:liver}, ContourDiff outperforms representative GAN-based (CycleGAN) and diffusion-based (UNSB) methods in liver translation, achieving superior performance in both segmentation accuracy and edge alignment, which demonstrates its capability for soft-tissue translation.

    \subsection{Evaluation on Higher Bits Normalization}
\begin{figure}[!ht]
        \centering
        \includegraphics[width=1\columnwidth]{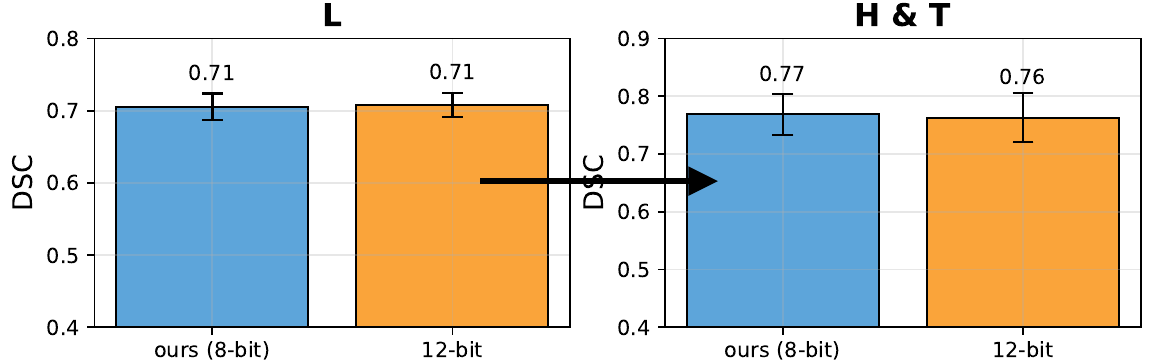}
        \caption{\textbf{Model performance with higher bit normalization}. 8-bit is the current setting.}
        \label{fig:robustness_vs_range}
    \end{figure}

    To evaluate whether higher bit normalization affects performance, we compared our current 8-bit normalization (i.e., 0-255) with 12-bit normalization (i.e., 0-4095). As shown in Figure \ref{fig:robustness_vs_range}, the results show no substantial performance differences between the two settings.

    \subsection{Efficiency Evaluation}
    To assess the practicality and feasibility of clinical deployment, we report the runtime and the peak memory usage of ContourDiff as shown in Table \ref{tab:result_efficiency}:

    \begin{table}[!h]
\centering
\resizebox{1\columnwidth}{!}{%
\begin{tabular}{l|c}
\toprule
Metric/Equipment & Results \\
\midrule
GPU & 1 NVIDIA RTX A6000 \\
Training Batch Size & 4 \\
Peak Training GPU Usage & 12.06 GB \\
\# Candidates for initial slices & 16 \\
Peak Testing GPU Usage & 10.40 GB \\
\# DDIM Steps & 50 \\
Inference Time per 2D Slice ($n$ parallel group) & 1.592 s $/ n$ \\
Inference Time per 3D Volume ($n$ parallel group) &  210.789 s $/ n$ \\
Ave. \# slices per 3D Volume &  132.4 \\
Total Training Time for ContourDiff & $\sim$ 5 hrs \\
Total Training Time for Segmentation & L: $\sim$ 20 min; H\&T: $\sim$ 1 hr \\
\bottomrule
\end{tabular}%
}
\caption{Efficiency assessment of ContourDiff.}
\label{tab:result_efficiency}
\end{table}

\section{Conclusion and Future Work}
\label{sec:conclusion}
    In this paper, we introduce a novel framework, ContourDiff with SCGD, to preserve anatomical fidelity in unpaired image translation. Our method constrains the generated images in the output domain to align with the anatomical contour of images from the input domain. Both quantitative and qualitative results on medical datasets show that ContourDiff (with/without SCGD) significantly outperforms multiple existing image translation methods in maintaining anatomical structures. Furthermore, we demonstrated the zero-shot capability of ContourDiff by translating T2-weighted MRI to T1-weighted MRI without any retraining.

    As a direction for future work, the practical deployment of ContourDiff may further benefit from automatic threshold selection based on simple image statistics (e.g., percentile-based methods) and from including a lightweight contour-refinement network, thereby further reducing manual tuning. We also note that susceptibility-related MRI distortions can carry clinically relevant information, thus, strictly enforcing pixel-wise contour alignment could risk suppressing such meaningful distortions. Future research will explore distortion-aware guidance to balance structural fidelity with the preservation of diagnostically relevant geometric information. In addition, while ContourDiff reliably preserves common anatomical structures such as bone, fat, muscle, and major organs, applications requiring finer tissue characteristics may benefit from integrating additional input-domain information or a hybrid conditioning strategy. Moreover, incorporating real multi-contrast or multi-echo MRI data and evaluating performance under varying acquisition conditions would further enhance the generalizability of the proposed methods across diverse clinical settings.



%
\ethics{The research protocol was approved by the Duke Health System Institutional Review Board (IRB) with ethical standards for research and manuscript preparation, adhering to all relevant laws and regulations concerning the treatment of human subjects and animals.}

\coi{We declare we do not have any conflicts of interest.}

\data{The external dataset analyzed in this study is publicly available (\cite{Wasserthal_2023,vandergraaf2023lumbar,ji2022amos}). The internal dataset, however, is not currently available due to an extensive de-identification procedures and institutional review board requirements. The code have been made publicly available at:~\url{https://github.com/mazurowski-lab/ContourDiff}}.

\bibliography{sample}

@inproceedings{konz2024anatomically,
  title={Anatomically-controllable medical image generation with segmentation-guided diffusion models},
  author={Konz, Nicholas and Chen, Yuwen and Dong, Haoyu and Mazurowski, Maciej A},
  booktitle={International Conference on Medical Image Computing and Computer-Assisted Intervention},
  pages={88--98},
  year={2024},
  organization={Springer}
}

@inproceedings{yang2019unsupervised,
  title={Unsupervised domain adaptation via disentangled representations: Application to cross-modality liver segmentation},
  author={Yang, Junlin and Dvornek, Nicha C and Zhang, Fan and Chapiro, Julius and Lin, MingDe and Duncan, James S},
  booktitle={Medical Image Computing and Computer Assisted Intervention--MICCAI 2019: 22nd International Conference, Shenzhen, China, October 13--17, 2019, Proceedings, Part II 22},
  pages={255--263},
  year={2019},
  organization={Springer}
}

@inproceedings{modanwal2020mri,
  title={MRI image harmonization using cycle-consistent generative adversarial network},
  author={Modanwal, Gourav and Vellal, Adithya and Buda, Mateusz and Mazurowski, Maciej A},
  booktitle={Medical Imaging 2020: Computer-Aided Diagnosis},
  volume={11314},
  pages={259--264},
  year={2020},
  organization={SPIE}
}

@inproceedings{liu2021style,
  title={Style transfer using generative adversarial networks for multi-site mri harmonization},
  author={Liu, Mengting and Maiti, Piyush and Thomopoulos, Sophia and Zhu, Alyssa and Chai, Yaqiong and Kim, Hosung and Jahanshad, Neda},
  booktitle={Medical Image Computing and Computer Assisted Intervention--MICCAI 2021: 24th International Conference, Strasbourg, France, September 27--October 1, 2021, Proceedings, Part III 24},
  pages={313--322},
  year={2021},
  organization={Springer}
}

@inproceedings{beizaee2023harmonizing,
  title={Harmonizing Flows: Unsupervised MR harmonization based on normalizing flows},
  author={Beizaee, Farzad and Desrosiers, Christian and Lodygensky, Gregory A and Dolz, Jose},
  booktitle={International Conference on Information Processing in Medical Imaging},
  pages={347--359},
  year={2023},
  organization={Springer}
}

@article{li2023zero,
  title={Zero-shot Medical Image Translation via Frequency-Guided Diffusion Models},
  author={Li, Yunxiang and Shao, Hua-Chieh and Liang, Xiao and Chen, Liyuan and Li, Ruiqi and Jiang, Steve and Wang, Jing and Zhang, You},
  journal={arXiv preprint arXiv:2304.02742},
  year={2023}
}

@inproceedings{cyclegan,
  title={Unpaired image-to-image translation using cycle-consistent adversarial networks},
  author={Zhu, Jun-Yan and Park, Taesung and Isola, Phillip and Efros, Alexei A},
  booktitle={Proceedings of the IEEE international conference on computer vision},
  pages={2223--2232},
  year={2017}
}

@inproceedings{controlnet,
  title={Adding conditional control to text-to-image diffusion models},
  author={Zhang, Lvmin and Rao, Anyi and Agrawala, Maneesh},
  booktitle={Proceedings of the IEEE/CVF International Conference on Computer Vision},
  pages={3836--3847},
  year={2023}
}

@article{ho2020denoising,
  title={Denoising diffusion probabilistic models},
  author={Ho, Jonathan and Jain, Ajay and Abbeel, Pieter},
  journal={Advances in neural information processing systems},
  volume={33},
  pages={6840--6851},
  year={2020}
}

@inproceedings{nichol2021improved,
  title={Improved denoising diffusion probabilistic models},
  author={Nichol, Alexander Quinn and Dhariwal, Prafulla},
  booktitle={International Conference on Machine Learning},
  pages={8162--8171},
  year={2021},
  organization={PMLR}
}

@inproceedings{unet,
  title={U-net: Convolutional networks for biomedical image segmentation},
  author={Ronneberger, Olaf and Fischer, Philipp and Brox, Thomas},
  booktitle={Medical Image Computing and Computer-Assisted Intervention--MICCAI 2015: 18th International Conference, Munich, Germany, October 5-9, 2015, Proceedings, Part III 18},
  pages={234--241},
  year={2015},
  organization={Springer}
}

@article{fid,
  title={Gans trained by a two time-scale update rule converge to a local nash equilibrium},
  author={Heusel, Martin and Ramsauer, Hubert and Unterthiner, Thomas and Nessler, Bernhard and Hochreiter, Sepp},
  journal={Advances in neural information processing systems},
  volume={30},
  year={2017}
}

@article{Wasserthal_2023,
   title={TotalSegmentator: Robust Segmentation of 104 Anatomic Structures in CT Images},
   volume={5},
   ISSN={2638-6100},
   url={http://dx.doi.org/10.1148/ryai.230024},
   DOI={10.1148/ryai.230024},
   number={5},
   journal={Radiology: Artificial Intelligence},
   publisher={Radiological Society of North America (RSNA)},
   author={Wasserthal, Jakob and Breit, Hanns-Christian and Meyer, Manfred T. and Pradella, Maurice and Hinck, Daniel and Sauter, Alexander W. and Heye, Tobias and Boll, Daniel T. and Cyriac, Joshy and Yang, Shan and Bach, Michael and Segeroth, Martin},
   year={2023},
   month=sep }

@misc{vandergraaf2023lumbar,
      title={Lumbar spine segmentation in MR images: a dataset and a public benchmark}, 
      author={Jasper W. van der Graaf and Miranda L. van Hooff and Constantinus F. M. Buckens and Matthieu Rutten and Job L. C. van Susante and Robert Jan Kroeze and Marinus de Kleuver and Bram van Ginneken and Nikolas Lessmann},
      year={2023},
      eprint={2306.12217},
      archivePrefix={arXiv},
      primaryClass={eess.IV}
}

@inproceedings{cao2022swin,
  title={Swin-unet: Unet-like pure transformer for medical image segmentation},
  author={Cao, Hu and Wang, Yueyue and Chen, Joy and Jiang, Dongsheng and Zhang, Xiaopeng and Tian, Qi and Wang, Manning},
  booktitle={European conference on computer vision},
  pages={205--218},
  year={2022},
  organization={Springer}
}

@article{synsegnet,
   title={SynSeg-Net: Synthetic Segmentation Without Target Modality Ground Truth},
   volume={38},
   ISSN={1558-254X},
   url={http://dx.doi.org/10.1109/TMI.2018.2876633},
   DOI={10.1109/tmi.2018.2876633},
   number={4},
   journal={IEEE Transactions on Medical Imaging},
   publisher={Institute of Electrical and Electronics Engineers (IEEE)},
   author={Huo, Yuankai and Xu, Zhoubing and Moon, Hyeonsoo and Bao, Shunxing and Assad, Albert and Moyo, Tamara K. and Savona, Michael R. and Abramson, Richard G. and Landman, Bennett A.},
   year={2019},
   month=apr, pages={1016–1025} }

@misc{cycada,
      title={CyCADA: Cycle-Consistent Adversarial Domain Adaptation}, 
      author={Judy Hoffman and Eric Tzeng and Taesung Park and Jun-Yan Zhu and Phillip Isola and Kate Saenko and Alexei A. Efros and Trevor Darrell},
      year={2017},
      eprint={1711.03213},
      archivePrefix={arXiv},
      primaryClass={cs.CV}
}

@inproceedings{
song2021ddim,
title={Denoising Diffusion Implicit Models},
author={Jiaming Song and Chenlin Meng and Stefano Ermon},
booktitle={International Conference on Learning Representations},
year={2021},
url={https://openreview.net/forum?id=St1giarCHLP}
}

@article{van_der_Walt_2014,
   title={scikit-image: image processing in Python},
   volume={2},
   ISSN={2167-8359},
   url={http://dx.doi.org/10.7717/peerj.453},
   DOI={10.7717/peerj.453},
   journal={PeerJ},
   publisher={PeerJ},
   author={van der Walt, Stéfan and Schönberger, Johannes L. and Nunez-Iglesias, Juan and Boulogne, François and Warner, Joshua D. and Yager, Neil and Gouillart, Emmanuelle and Yu, Tony},
   year={2014},
   month=jun, pages={e453} }

@inproceedings{maskgan,
  title={Structure-Preserving Synthesis: MaskGAN for Unpaired MR-CT Translation},
  author={Phan, Vu Minh Hieu and Liao, Zhibin and Verjans, Johan W and To, Minh-Son},
  booktitle={International Conference on Medical Image Computing and Computer-Assisted Intervention},
  pages={56--65},
  year={2023},
  organization={Springer}
}

@article{armanious2020medgan,
  title={MedGAN: Medical image translation using GANs},
  author={Armanious, Karim and Jiang, Chenming and Fischer, Marc and K{\"u}stner, Thomas and Hepp, Tobias and Nikolaou, Konstantin and Gatidis, Sergios and Yang, Bin},
  journal={Computerized medical imaging and graphics},
  volume={79},
  pages={101684},
  year={2020},
  publisher={Elsevier}
}

@article{chen2023deep,
  title={Deep learning based unpaired image-to-image translation applications for medical physics: a systematic review},
  author={Chen, Junhua and Chen, Shenlun and Wee, Leonard and Dekker, Andre and Bermejo, Inigo},
  journal={Physics in Medicine \& Biology},
  year={2023},
  publisher={IOP Publishing}
}

@article{li2020magnetic,
  title={Magnetic resonance image (MRI) synthesis from brain computed tomography (CT) images based on deep learning methods for magnetic resonance (MR)-guided radiotherapy},
  author={Li, Wen and Li, Yafen and Qin, Wenjian and Liang, Xiaokun and Xu, Jianyang and Xiong, Jing and Xie, Yaoqin},
  journal={Quantitative imaging in medicine and surgery},
  volume={10},
  number={6},
  pages={1223},
  year={2020},
  publisher={AME Publications}
}

@article{rossi2021comparison,
  title={Comparison of supervised and unsupervised approaches for the generation of synthetic CT from cone-beam CT},
  author={Rossi, Matteo and Cerveri, Pietro},
  journal={Diagnostics},
  volume={11},
  number={8},
  pages={1435},
  year={2021},
  publisher={MDPI}
}

@article{canny1986computational,
  title={A computational approach to edge detection},
  author={Canny, John},
  journal={IEEE Transactions on pattern analysis and machine intelligence},
  number={6},
  pages={679--698},
  year={1986},
  publisher={Ieee}
}

@article{ozbey2023unsupervised,
  title={Unsupervised medical image translation with adversarial diffusion models},
  author={{\"O}zbey, Muzaffer and Dalmaz, Onat and Dar, Salman UH and Bedel, Hasan A and {\"O}zturk, {\c{S}}aban and G{\"u}ng{\"o}r, Alper and {\c{C}}ukur, Tolga},
  journal={IEEE Transactions on Medical Imaging},
  year={2023},
  publisher={IEEE}
}

@inproceedings{armanious2019unsupervised,
  title={Unsupervised medical image translation using cycle-MedGAN},
  author={Armanious, Karim and Jiang, Chenming and Abdulatif, Sherif and K{\"u}stner, Thomas and Gatidis, Sergios and Yang, Bin},
  booktitle={2019 27th European signal processing conference (EUSIPCO)},
  pages={1--5},
  year={2019},
  organization={IEEE}
}

@article{zhou2023gan,
  title={GAN review: Models and medical image fusion applications},
  author={Zhou, Tao and Li, Qi and Lu, Huiling and Cheng, Qianru and Zhang, Xiangxiang},
  journal={Information Fusion},
  volume={91},
  pages={134--148},
  year={2023},
  publisher={Elsevier}
}

@article{goodfellow2020generative,
  title={Generative adversarial networks},
  author={Goodfellow, Ian and Pouget-Abadie, Jean and Mirza, Mehdi and Xu, Bing and Warde-Farley, David and Ozair, Sherjil and Courville, Aaron and Bengio, Yoshua},
  journal={Communications of the ACM},
  volume={63},
  number={11},
  pages={139--144},
  year={2020},
  publisher={ACM New York, NY, USA}
}

@inproceedings{li2023bbdm,
  title={Bbdm: Image-to-image translation with brownian bridge diffusion models},
  author={Li, Bo and Xue, Kaitao and Liu, Bin and Lai, Yu-Kun},
  booktitle={Proceedings of the IEEE/CVF conference on computer vision and pattern Recognition},
  pages={1952--1961},
  year={2023}
}

@inproceedings{kim2024adaptive,
  title={Adaptive latent diffusion model for 3d medical image to image translation: Multi-modal magnetic resonance imaging study},
  author={Kim, Jonghun and Park, Hyunjin},
  booktitle={Proceedings of the IEEE/CVF Winter Conference on Applications of Computer Vision},
  pages={7604--7613},
  year={2024}
}

@article{batzolis2021conditional,
  title={Conditional image generation with score-based diffusion models},
  author={Batzolis, Georgios and Stanczuk, Jan and Sch{\"o}nlieb, Carola-Bibiane and Etmann, Christian},
  journal={arXiv preprint arXiv:2111.13606},
  year={2021}
}

@inproceedings{rombach2022high,
  title={High-resolution image synthesis with latent diffusion models},
  author={Rombach, Robin and Blattmann, Andreas and Lorenz, Dominik and Esser, Patrick and Ommer, Bj{\"o}rn},
  booktitle={Proceedings of the IEEE/CVF conference on computer vision and pattern recognition},
  pages={10684--10695},
  year={2022}
}

@inproceedings{isola2017image,
  title={Image-to-image translation with conditional adversarial networks},
  author={Isola, Phillip and Zhu, Jun-Yan and Zhou, Tinghui and Efros, Alexei A},
  booktitle={Proceedings of the IEEE conference on computer vision and pattern recognition},
  pages={1125--1134},
  year={2017}
}

@inproceedings{wang2018high,
  title={High-resolution image synthesis and semantic manipulation with conditional gans},
  author={Wang, Ting-Chun and Liu, Ming-Yu and Zhu, Jun-Yan and Tao, Andrew and Kautz, Jan and Catanzaro, Bryan},
  booktitle={Proceedings of the IEEE conference on computer vision and pattern recognition},
  pages={8798--8807},
  year={2018}
}

@inproceedings{munit,
  title={Multimodal unsupervised image-to-image translation},
  author={Huang, Xun and Liu, Ming-Yu and Belongie, Serge and Kautz, Jan},
  booktitle={Proceedings of the European conference on computer vision (ECCV)},
  pages={172--189},
  year={2018}
}

@inproceedings{gcgan,
  title={Geometry-consistent generative adversarial networks for one-sided unsupervised domain mapping},
  author={Fu, Huan and Gong, Mingming and Wang, Chaohui and Batmanghelich, Kayhan and Zhang, Kun and Tao, Dacheng},
  booktitle={Proceedings of the IEEE/CVF conference on computer vision and pattern recognition},
  pages={2427--2436},
  year={2019}
}

@inproceedings{cut,
  title={Contrastive learning for unpaired image-to-image translation},
  author={Park, Taesung and Efros, Alexei A and Zhang, Richard and Zhu, Jun-Yan},
  booktitle={Computer Vision--ECCV 2020: 16th European Conference, Glasgow, UK, August 23--28, 2020, Proceedings, Part IX 16},
  pages={319--345},
  year={2020},
  organization={Springer}
}

@article{unsb,
  title={Unpaired Image-to-Image Translation via Neural Schr\"{o}dinger Bridge},
  author={Kim, Beomsu and Kwon, Gihyun and Kim, Kwanyoung and Ye, Jong Chul},
  journal={arXiv preprint arXiv:2305.15086},
  year={2023}
}

@article{croitoru2023diffusion,
  title={Diffusion models in vision: A survey},
  author={Croitoru, Florinel-Alin and Hondru, Vlad and Ionescu, Radu Tudor and Shah, Mubarak},
  journal={IEEE Transactions on Pattern Analysis and Machine Intelligence},
  volume={45},
  number={9},
  pages={10850--10869},
  year={2023},
  publisher={IEEE}
}

@inproceedings{zhang2018translating,
  title={Translating and segmenting multimodal medical volumes with cycle-and shape-consistency generative adversarial network},
  author={Zhang, Zizhao and Yang, Lin and Zheng, Yefeng},
  booktitle={Proceedings of the IEEE conference on computer vision and pattern Recognition},
  pages={9242--9251},
  year={2018}
}

@article{wang2024mutual,
  title={Mutual information guided diffusion for zero-shot cross-modality medical image translation},
  author={Wang, Zihao and Yang, Yingyu and Chen, Yuzhou and Yuan, Tingting and Sermesant, Maxime and Delingette, Herv{\'e} and Wu, Ona},
  journal={IEEE Transactions on Medical Imaging},
  year={2024},
  publisher={IEEE}
}

@inproceedings{durrer2024diffusion,
  title={Diffusion Models for Contrast Harmonization of Magnetic Resonance Images},
  author={Durrer, Alicia and Wolleb, Julia and Bieder, Florentin and Sinnecker, Tim and Weigel, Matthias and Sandkuehler, Robin and Granziera, Cristina and Yaldizli, {\"O}zg{\"u}r and Cattin, Philippe C},
  booktitle={Medical Imaging with Deep Learning},
  pages={526--551},
  year={2024}
}

@article{muller2023multimodal,
  title={A multimodal comparison of latent denoising diffusion probabilistic models and generative adversarial networks for medical image synthesis},
  author={M{\"u}ller-Franzes, Gustav and Niehues, Jan Moritz and Khader, Firas and Arasteh, Soroosh Tayebi and Haarburger, Christoph and Kuhl, Christiane and Wang, Tianci and Han, Tianyu and Nolte, Teresa and Nebelung, Sven and others},
  journal={Scientific Reports},
  volume={13},
  number={1},
  pages={12098},
  year={2023},
  publisher={Nature Publishing Group UK London}
}

@article{saharia2022image,
  title={Image super-resolution via iterative refinement},
  author={Saharia, Chitwan and Ho, Jonathan and Chan, William and Salimans, Tim and Fleet, David J and Norouzi, Mohammad},
  journal={IEEE transactions on pattern analysis and machine intelligence},
  volume={45},
  number={4},
  pages={4713--4726},
  year={2022},
  publisher={IEEE}
}

@inproceedings{gao2023implicit,
  title={Implicit diffusion models for continuous super-resolution},
  author={Gao, Sicheng and Liu, Xuhui and Zeng, Bohan and Xu, Sheng and Li, Yanjing and Luo, Xiaoyan and Liu, Jianzhuang and Zhen, Xiantong and Zhang, Baochang},
  booktitle={Proceedings of the IEEE/CVF conference on computer vision and pattern recognition},
  pages={10021--10030},
  year={2023}
}

@inproceedings{corneanu2024latentpaint,
  title={Latentpaint: Image inpainting in latent space with diffusion models},
  author={Corneanu, Ciprian and Gadde, Raghudeep and Martinez, Aleix M},
  booktitle={Proceedings of the IEEE/CVF Winter Conference on Applications of Computer Vision},
  pages={4334--4343},
  year={2024}
}

@article{tan2022semantic,
  title={Semantic diffusion network for semantic segmentation},
  author={Tan, Haoru and Wu, Sitong and Pi, Jimin},
  journal={Advances in Neural Information Processing Systems},
  volume={35},
  pages={8702--8716},
  year={2022}
}

@article{baranchuk2021label,
  title={Label-efficient semantic segmentation with diffusion models},
  author={Baranchuk, Dmitry and Rubachev, Ivan and Voynov, Andrey and Khrulkov, Valentin and Babenko, Artem},
  journal={arXiv preprint arXiv:2112.03126},
  year={2021}
}

@article{chen2023moco,
  title={MoCo-Transfer: Investigating out-of-distribution contrastive learning for limited-data domains},
  author={Chen, Yuwen and Zhou, Helen and Lipton, Zachary C},
  journal={arXiv preprint arXiv:2311.09401},
  year={2023}
}

@article{kong2021breaking,
  title={Breaking the dilemma of medical image-to-image translation},
  author={Kong, Lingke and Lian, Chenyu and Huang, Detian and Hu, Yanle and Zhou, Qichao and others},
  journal={Advances in Neural Information Processing Systems},
  volume={34},
  pages={1964--1978},
  year={2021}
}

@article{uzunova2020memory,
  title={Memory-efficient GAN-based domain translation of high resolution 3D medical images},
  author={Uzunova, Hristina and Ehrhardt, Jan and Handels, Heinz},
  journal={Computerized Medical Imaging and Graphics},
  volume={86},
  pages={101801},
  year={2020},
  publisher={Elsevier}
}

@article{kid,
  title={Demystifying mmd gans},
  author={Bi{\'n}kowski, Miko{\l}aj and Sutherland, Danica J and Arbel, Michael and Gretton, Arthur},
  journal={arXiv preprint arXiv:1801.01401},
  year={2018}
}

@article{ji2022amos,
  title={Amos: A large-scale abdominal multi-organ benchmark for versatile medical image segmentation},
  author={Ji, Yuanfeng and Bai, Haotian and Ge, Chongjian and Yang, Jie and Zhu, Ye and Zhang, Ruimao and Li, Zhen and Zhanng, Lingyan and Ma, Wanling and Wan, Xiang and others},
  journal={Advances in neural information processing systems},
  volume={35},
  pages={36722--36732},
  year={2022}
}

@article{song2020denoising,
  title={Denoising diffusion implicit models},
  author={Song, Jiaming and Meng, Chenlin and Ermon, Stefano},
  journal={arXiv preprint arXiv:2010.02502},
  year={2020}
}

@article{ma2024segment,
  title={Segment anything in medical images},
  author={Ma, Jun and He, Yuting and Li, Feifei and Han, Lin and You, Chenyu and Wang, Bo},
  journal={Nature Communications},
  volume={15},
  number={1},
  pages={654},
  year={2024},
  publisher={Nature Publishing Group UK London}
}

@inproceedings{lyu2024superpixel,
  title={Superpixel-Guided Segment Anything Model for Liver Tumor Segmentation with Couinaud Segment Prompt},
  author={Lyu, Fei and Xu, Jingwen and Zhu, Ye and Wong, Grace Lai-Hung and Yuen, Pong C},
  booktitle={International Conference on Medical Image Computing and Computer-Assisted Intervention},
  pages={678--688},
  year={2024},
  organization={Springer}
}

@article{mazurowski2023segment,
  title={Segment anything model for medical image analysis: an experimental study},
  author={Mazurowski, Maciej A and Dong, Haoyu and Gu, Hanxue and Yang, Jichen and Konz, Nicholas and Zhang, Yixin},
  journal={Medical Image Analysis},
  volume={89},
  pages={102918},
  year={2023},
  publisher={Elsevier}
}





\end{document}